\documentclass[notitlepage,floatfix,superscriptaddress,nofootinbib,tightenlines,preprintnumbers,twocolumn]{revtex4-2}
\usepackage{amsmath,verbatim,latexsym,amssymb,indentfirst,mathrsfs,mathtools,amsthm,bbm,bm,hyperref,url,cancel,subcaption,enumitem}
\usepackage[font=small,labelfont=bf,format=plain,justification=raggedright,singlelinecheck=false]{caption}
\usepackage{verbatim,indentfirst}
\usepackage{siunitx}
\usepackage[title,titletoc]{appendix}
\hypersetup{nolinks=true}

\usepackage{graphicx}
\usepackage{epstopdf}
\newcommand{\caphead}[1]{{\bf #1}}

\renewcommand{\thesection}{\Roman{section}}
\renewcommand{\thesubsection}{\Roman{section} \Alph{subsection}}
\renewcommand{\thesubsubsection}{\Roman{section} \Alph{subsection} \arabic{subsubsection}}
\makeatletter
\def\p@subsection{}
\makeatother
\makeatletter
\def\p@subsubsection{}
\makeatother

\newtheorem{theorem}{Theorem}

\makeatletter
\newcommand\footnoteref[1]{\protected@xdef\@thefnmark{\ref{#1}}\@footnotemark}
\makeatother

\usepackage{soul}

\usepackage{color}
\usepackage[dvipsnames]{xcolor}
\usepackage[normalem]{ulem}

\newcommand{\Heis}{{\rm Heis}}

\newcommand{\numSteps}{{N_{\rm T}}}
\newcommand{\NATS}{{\rm NATS}}

\newcommand{\final}{{\rm f}}
\newcommand{\can}{{\rm can}}
\newcommand{\GC}{{\rm GC}}
\newcommand{\exact}{{\rm exact}}  
\newcommand{\depol}{{\rm depol}}  
\newcommand{\trans}{\tau}  
\newcommand{\Trans}{\mathcal{T}}  
\newcommand{\avg}{{\rm avg}}
\newcommand{\dist}{\mathcal{D}_{\rm tr}}

\newcommand{\tot}{ {\rm tot} }
\newcommand{\Tr}{{\rm Tr}}   
\def\id{\mathbbm{1}}   

\newcommand{\Sys}{\mathcal{S}}  
\newcommand{\Sites}{N}  

\newcommand{\1}{ {(1)} }

\newcommand{\JParen}{ {(j)} }
\newcommand{\KParen}{ {(k)} }
\newcommand{\LParen}{ \bm{(} }
\newcommand{\RParen}{ \bm{)} }

\newcommand*{\Set}[1]{\left\{  #1  \right\}}

	\definecolor{blue(pigment)}{rgb}{0.2, 0.2, 0.6}
 
\renewcommand\th{ {\rm th} }

\newcommand*{\bra}[1]{\langle #1\rvert}
\newcommand*{\ket}[1]{\lvert #1 \rangle}

\newcommand*{\ketbra}[2]{\lvert #1 \rangle\!\langle #2 \rvert}
\newcommand*{\expval}[1]{\left\langle  #1  \right\rangle}

\hypersetup{final}

\begin{document}

\title{Experimental observation of thermalization with noncommuting charges}
\author{Florian Kranzl}
\email{The first two coauthors contributed equally.}
\affiliation{Institut f\"ur Quantenoptik und Quanteninformation, \"Osterreichische Akademie der Wissenschaften,
Technikerstra\ss{}e 21a, 6020 Innsbruck, Austria}
\affiliation{Institut f\"ur Experimentalphysik, Universit\"at Innsbruck, Technikerstra\ss{}e 25, 6020 Innsbruck, Austria}
\author{Aleksander~Lasek}
\email{The first two coauthors contributed equally.}
\affiliation{Joint Center for Quantum Information and Computer Science, NIST and University of Maryland, College Park, MD 20742, USA}
\author{Manoj K. Joshi}
\affiliation{Institut f\"ur Quantenoptik und Quanteninformation, \"Osterreichische Akademie der Wissenschaften,
Technikerstra\ss{}e 21a, 6020 Innsbruck, Austria}
\author{Amir Kalev}
\affiliation{Information Sciences Institute, University of Southern California, Arlington, VA 22203, USA}
\author{Rainer Blatt}
\affiliation{Institut f\"ur Quantenoptik und Quanteninformation, \"Osterreichische Akademie der Wissenschaften,
Technikerstra\ss{}e 21a, 6020 Innsbruck, Austria}
\affiliation{Institut f\"ur Experimentalphysik, Universit\"at Innsbruck, Technikerstra\ss{}e 25, 6020 Innsbruck, Austria}
\author{Christian F. Roos}
\email{christian.roos@uibk.ac.at}
\affiliation{Institut f\"ur Quantenoptik und Quanteninformation, \"Osterreichische Akademie der Wissenschaften,
Technikerstra\ss{}e 21a, 6020 Innsbruck, Austria}
\affiliation{Institut f\"ur Experimentalphysik, Universit\"at Innsbruck, Technikerstra\ss{}e 25, 6020 Innsbruck, Austria}
\author{Nicole~Yunger~Halpern}
\email{nicoleyh@umd.edu}
\affiliation{Joint Center for Quantum Information and Computer Science, NIST and University of Maryland, College Park, MD 20742, USA}
\affiliation{Institute for Physical Science and Technology, University of Maryland, College Park, MD 20742, USA}
\affiliation{ITAMP, Harvard-Smithsonian Center for Astrophysics, Cambridge, MA 02138, USA}
\affiliation{Department of Physics, Harvard University, Cambridge, MA 02138, USA}
\date{\today}

%
%
\begin{abstract} 
Quantum simulators have recently enabled experimental observations of quantum many-body systems’ internal thermalization. Often, the global energy and particle number are conserved, and the system is prepared with a well-defined particle number---in a microcanonical subspace. However, quantum evolution can also conserve quantities, or charges, that fail to commute with each other. Noncommuting charges have recently emerged as a subfield at the intersection of quantum thermodynamics and quantum information. Until now, this subfield has remained theoretical. We initiate the experimental testing of its predictions, with a trapped-ion simulator. We prepare 6--21 spins in an approximate microcanonical subspace, a generalization of the microcanonical subspace for accommodating noncommuting charges, which cannot necessarily have well-defined nontrivial values simultaneously. We simulate a Heisenberg evolution using laser-induced entangling interactions and collective spin rotations. The noncommuting charges are the three spin components. We find that small subsystems equilibrate to near a recently predicted non-Abelian thermal state. This work bridges quantum many-body simulators to the quantum thermodynamics of noncommuting charges, whose predictions can now be tested.
\end{abstract}

{\let\newpage\relax\maketitle}

%
%
%

Thermalization aims the arrow of time, yet has traditionally been understood through the lens of classical systems. Understanding quantum thermalization is therefore of fundamental importance.
Quantum-simulator experiments have recently elucidated how
closed quantum many-body systems thermalize internally~\cite{Kaufman_16_Quantum, Neill_16_Ergodic, Clos2016thermalization, Zhou_22_Thermalization}.
Typically, the evolutions conserve no quantities
(as in gate-based evolutions) or 
conserve energy and particle number (in analog quantum simulators). The conserved quantities, called \emph{charges}, are represented by Hermitian operators $Q_{\gamma = 1, 2, \ldots, c}$.
The operators are usually assumed implicitly to commute with each other, as do the commonly conserved Hamiltonian and particle-number operator.
Yet noncommuting operators underlie quantum physics
from uncertainty relations to measurement disturbance.
What happens if thermodynamic charges fail to commute with each other? This question recently swept across quantum thermodynamics~\cite{Lostaglio_14_Masters,NYH_18_Beyond, Guryanova_16_Thermodynamics,NYH_16_Microcanonical,Lostaglio_17_Thermodynamic, Sparaciari_18_First, Khanian_20_From, Khanian_20_Resource, Gour_18_Quantum, Manzano_22_Non, Popescu_18_Quantum, Popescu_19_Reference, Lostaglio_14_Masters, NYH_16_Microcanonical, Ito_18_Optimal, Bera_19_Thermo, Mur_Petit_18_Revealing, Manzano_18_Squeezed,  NYH_20_Noncommuting, Manzano_20_Hybrid, Fukai_20_Noncommutative, Mur_Petit_19_Fluctuations, Scandi_18_Thermodynamic, Manzano_18_Squeezed, Sparaciari_18_First, Mur_Petit_18_Revealing, Boes_18_Statistical, Ito_18_Optimal, Mitsuhashi_22_Characterizing, Croucher_18_Information, Vaccaro_11_Information, Wright_18_Quantum,Zhang_20_Stationary,Zhang_20_Stationary,Medenjak_20_Isolated,NYH_22_How, Croucher_21_Memory,Marvian_21_Qudit,Marvian_22_Rotationally,Ducuara_22_Quantum,Murthy_22_Non,Majidy_22_Non} 
and infiltrated many-body theory~\cite{Manzano_18_Squeezed,Mur_Petit_18_Revealing,Mur_Petit_19_Fluctuations,NYH_20_Noncommuting,Manzano_20_Hybrid,Manzano_22_Non,Fukai_20_Noncommutative,NYH_22_How,Murthy_22_Non,Corps_22_Theory,Majidy_22_Non}.
We initiate experimentation on thermalization in the presence of noncommuting charges.

A many-body system thermalizes internally as a small subsystem $\Sys$ approaches the appropriate thermal state, which depends on the charges.
The rest of the global system acts as an effective environment.
Arguments for the thermal state's form rely implicitly on the charges' commutation~\cite{Renes_16_Beyond,NYH_18_Beyond,NYH_16_Microcanonical,NYH_20_Noncommuting}.
For example, the eigenstate thermalization hypothesis explains the internal thermalization of quantum many-body systems governed by nondegenerate Hamiltonians~\cite{Deutsch_91_Quantum,Srednicki_94_Chaos,Rigol_08_Thermalization}; yet noncommuting charges imply energy degeneracies.
Therefore, whether $\Sys$ can even thermalize, if charges fail to commute with each other, is not obvious.

Information-theoretic arguments suggest that 
$\Sys$ equilibrates to near a state dubbed
the \emph{non-Abelian thermal state} (NATS)~\cite{Jaynes_57_Information_II,NYH_18_Beyond,NYH_16_Microcanonical,Guryanova_16_Thermodynamics,Lostaglio_17_Thermodynamic},
\begin{align}
   \label{eq_NATS}
   \rho_\NATS
   := \exp \left( - \beta \left[ H
   - \sum_{\gamma = 1}^c 
   \mu_\gamma Q_\gamma \right]  \right) / Z_\NATS \ .
\end{align}
$\beta$ denotes the inverse temperature,
$H$ denotes the Hamiltonian of $\Sys$,
the $\mu_\gamma$ denote effective chemical potentials,
the $Q_\gamma$ denote the $c$ non-energy charges of $\Sys$,
and the partition function $Z_\NATS$ normalizes the state.
States of the form~\eqref{eq_NATS} are called also \emph{generalized Gibbs ensembles}, especially if the charges commute and the global Hamiltonian is integrable~\cite{Rigol_09_Breakdown,Rigol_07_Relaxation,Vidmar_16_Generalized}.
$\rho_\NATS$ has the exponential form typical of thermal states. Since the $Q_\gamma$ fail to commute, however, two common derivations of the thermal state's form break down~\cite{NYH_18_Beyond,NYH_16_Microcanonical}. For this reason, we distinguish $\rho_\NATS$ by the term \emph{non-Abelian}.
Arguments for Eq.~\eqref{eq_NATS} center on information theory; kinematics; and idealizations;
such as a very large system-and-environment composite~\cite{Jaynes_57_Information_II,NYH_16_Microcanonical,Guryanova_16_Thermodynamics,Lostaglio_17_Thermodynamic}.
Whether $\Sys$ thermalizes outside these idealizations, under realistic dynamics, has remained unclear.
Whether experimentalists can observe $\rho_\NATS$
has remained even unclearer: Experimental control is finite, so no quantum many-body system is truly closed. If many species of charge can leak out, many conservation laws can be violated.

Beyond these practicalities, to what extent noncommuting charges permit thermalization has been fundamentally unclear.
If just energy and particle number are conserved, then,
to thermalize $\Sys$, 
we prepare the global system in a \emph{microcanonical subspace}:
in a narrow energy window in a particle-number sector~\cite{Laundau_80_Statistical}.
If more charges are conserved, the microcanonical subspace is
a joint eigenspace shared by the $c$ global charges.
If the charges fail to commute,
they share no eigenbasis, so they may share no eigenspace:
No microcanonical subspace necessarily exists.
To accommodate noncommuting charges~\cite{NYH_16_Microcanonical},
microcanonical subspaces have been generalized to
\emph{approximate microcanonical (AMC) subspaces}.
In an AMC subspace, measuring any global charge has 
a high probability of yielding the expected value.
The uncertainty in the global charges' initial values
may generate uncertainty in 
the long-time state of $\Sys$:
$\Sys$ may remain farther from $\rho_\NATS$ 
than it would remain from the relevant thermal state if the charges commuted~\cite{NYH_20_Noncommuting}.
Furthermore, if charges fail to commute with each other, then (i) two derivations of the thermal state's form are invalid~\cite{NYH_16_Microcanonical,NYH_18_Beyond}; (ii) the Hamiltonian has degeneracies, which hinder arguments for thermalization~\cite{NYH_20_Noncommuting}; and (iii) the eigenstate thermalization hypothesis, one of the most widely used explanations of quantum many-body thermalization internally, breaks down~\cite{Murthy_22_Non}. Hence the extent to which noncommuting charges permit thermalization is unclear.

We experimentally observe thermalization to near $\rho_\NATS$, implementing the proposal in~\cite{NYH_20_Noncommuting}.
Our quantum simulator consists of 21 trapped ions. 
Two electronic states of each ion form a qubit. 
We initialise the qubits in an AMC subspace.
The evolution---an effective long-range Heisenberg coupling---conserves the global-spin components $S_{x,y,z}^\tot$.
We implement the evolution by interspersing 
a long-range Ising coupling with global rotations
and dynamical-decoupling sequences.
Trotterization of Heisenberg dynamics has been proposed theoretically~\cite{Viola_99_Universal,Jane_03_Simulation},
realized experimentally in toy examples \cite{Lanyon:2011,Salathe:2015}, and used very recently to explore many-body physics in ensembles of Rydberg atoms \cite{Geier2021Floquet,Scholl:2022};
we demonstrate its effectiveness in many-body experiments on trapped ions.
Two nearest-neighbor ions form the system of interest, 
the other ions forming an effective environment
(Fig.~\ref{fig_setup}).
We measure $\Sys$'s distance from $\rho_\NATS$,
finding significant thermalization 
on average over copies of $\Sys$~\cite{Jaynes_57_Information_II,NYH_18_Beyond,NYH_16_Microcanonical,Guryanova_16_Thermodynamics,Lostaglio_17_Thermodynamic}.
To begin to isolate the noncommutation's effects on thermalization,
we compare our experiment with an evolution that conserves just commuting charges: the Hamiltonian and $S_z^\tot$.
$\Sys$ remains farther from the thermal state
if the charges fail to commute.
This observation is consistent with the conjecture that
noncommuting charges hinder thermalization~\cite{NYH_16_Microcanonical}, as well as with the expectation that, in finite-size global systems, resistance to thermalization grows with the number of charges~\cite{Huse_21_Private,Monteiro_21_Quantum}.
Our experiment offers a particularly quantum counterpart to the landmark experiment~\cite{Langen_15_Experimental},
in which a hitherto-unobserved equilibrium state was observed
but the quantum physics of charges' noncommutation was left unexplored.
The present work opens the emerging subfield 
of noncommuting thermodynamic charges
to quantum many-body simulators.

\section{Experimental setup} 
\label{sec_Setup}

We begin by explaining the general experimental setup and protocol in Sec.~\ref{sec_Platform_Protocol}. Section~\ref{sec_Init_State} motivates and introduces our initial state.

\subsection{Platform and protocol}
\label{sec_Platform_Protocol}

\begin{figure*}[ht!]
\centering
    \includegraphics[width=180mm]{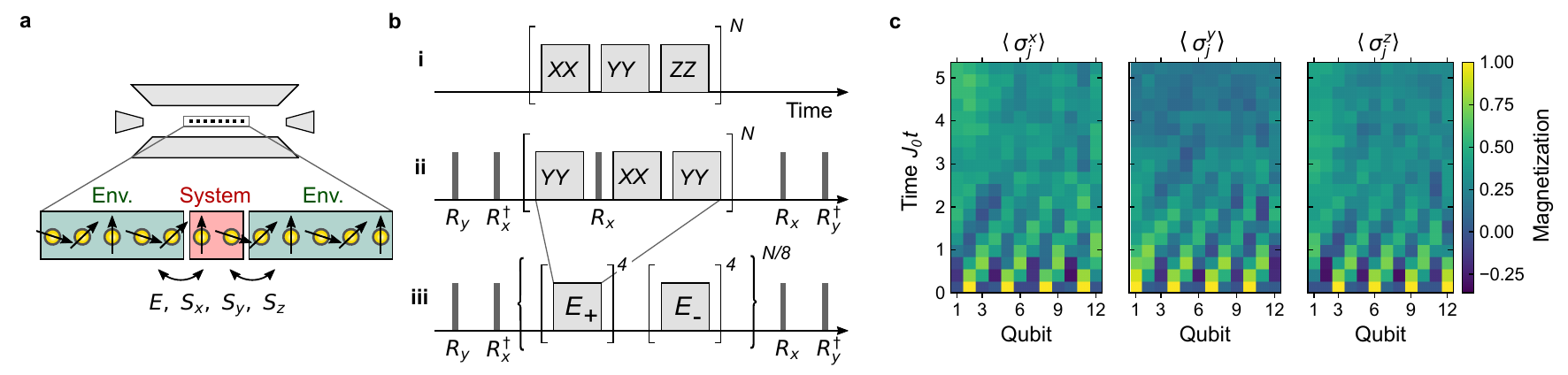}
    \caption{\caphead{Experimental setup and protocol:} 
    (a) A linear ion crystal of $\Sites \leq$ 21 qubits is trapped in a linear Paul trap. A small system exchanges charges (local instances of quantities that are conserved globally) with the surrounding environment: energy, $E$, and all the components of angular momentum. 
    (b) We Trotter-approximate the Heisenberg evolution by evolving the state under each of the Hamiltonian's three terms [Eq.~\eqref{eq_Heis}] consecutively, in short time steps. We experimentally realize two terms directly and generate the third term using resonant $\pi/2$-pulses ($R_x$ and $R_y$). This pulse sequence further protects the state against dephasing noise (ii). The Trotter sequence contains building blocks $E_\pm$. Alternating between them reduces pulse-length errors (iii). For further details, see 
    Sec.~\ref{sec_Methods}.
    (c) Observed evolution of the 12-qubit initial state, 
    $\ket{ y+, x+, z+ }^{\otimes 4}$,
    under the Trotter-approximated Heisenberg model
    (wherein $J_0=356$~rad/s and $\alpha=0.70$).
    To characterize the dynamics fully, we derive the spin-excitation-hopping rate in App.~\ref{app_Hopping}.
    }
    \label{fig_setup}
\end{figure*}

We perform the experiment on a trapped-ion quantum simulator~\cite{kranzl2022}. 
A linear string of $\Sites =$ 21
${}^{40}\mathrm{Ca}^{+}$ ions 
is confined in a linear Paul trap (Fig.~\ref{fig_setup}a). 
Charges' noncommutation is expected to influence many-body equilibration only in such mesoscale systems, as the correspondence principle dictates that systems grow classical as they grow large and charges' noncommutation is nonclassical~\cite{NYH_16_Microcanonical,NYH_20_Noncommuting}.
Let $\sigma_\gamma = \frac{2}{\hbar} S_\gamma$ denote the Pauli-$\gamma$ operator, for $\gamma = x, y, z$.
Let $\ket{\gamma \pm}$ denote the $\pm 1$ eigenstates of
$\sigma_\gamma$.
We denote by $\sigma_\gamma^\JParen$ 
the site-$j$ Pauli operators; 
and, by $\sigma_\gamma^\tot$, the whole-chain operators.
Each ion encodes a qubit in the Zeeman states 
$3^2\mathrm{D}_{5/2}$ and $4^2\mathrm{S}_{1/2}$,
of respective magnetic quantum numbers $m = 5/2$ and $1/2$.
We denote the states by $\ket{z+}$ and $\ket{z-}$.
Two nearest-neighbor qubits form
the small system of interest;
the remaining qubits form the environment.

We employ two types of coherent operations using a laser at 729 nm, which drives the quadrupole transition that connects the qubit states: 
(i) Denoting a rotated Pauli operator by
$\sigma_\phi^{(j)}= \cos \phi \, \sigma_x^{(j)} 
+ \sin \phi \, \sigma_y^{(j)}$,
we perform global qubit rotations
$U(\theta,\phi)
=\exp(-i \frac{\theta}{2} 
\sum_{j=1}^{\Sites} \sigma_\phi^{(j)})$.
(ii) The effective long-range $x$-type Ising Hamiltonian
\begin{align}
   \label{eq_IsingH}
   H_{xx} := \sum_{j<k} 
   \frac{J_0}{\left| j-k \right|^\alpha} \:
   \sigma_x^{(j)}  \sigma_x^{(k)},
\end{align}
entangles qubits.\footnote{
Long-range interactions are practical here because they internally thermalize the quantum many-body system rapidly.
The interaction time can therefore be short, giving the system little time to decohere. Short-range interactions have been shown numerically to induce thermalization to near the NATS~\cite{NYH_20_Noncommuting}.
}
We effect $H_{xx}$ by off-resonantly coupling to the lower and upper vibrational sideband transitions of the ion strings' transverse collective modes~\cite{Porras:2004}.
Combining these two ingredients, 
we Trotter-approximate the Heisenberg Hamiltonian
\begin{equation}
    \label{eq_Heis}
    H_\Heis
    := \sum_{j<k} \frac{J_0}{3 \left| j-k \right|^\alpha} \left( \sigma_x^{(j)} \sigma_x^{(k)}  
    + \sigma_y^{(j)} \sigma_y^{(k)} 
    + \sigma_z^{(j)} \sigma_z^{(k)} \right) \, ,
\end{equation}
as shown in Fig.~\ref{fig_setup}b and 
Sec.~\ref{sec_Methods}.
The $1/3$ appears because the Ising coupling~\eqref{eq_IsingH} is distributed across three directions ($x$, $y$, and $z$). 
We implement a $\sigma_y^\JParen \sigma_y^{(k)}$ 
coupling similarly, as described in 
Sec.~\ref{sec_Methods}.
The pulse sequence was designed to realize $H_{xx}$ while, via dynamical decoupling, mitigating dephasing and rotation errors.

At the beginning of each experimental trial, 
the ion string's transverse collective modes are cooled to near their motional ground state.
Then, we prepare the qubits in the product state described in 
the next subsection. 
We then evolve the global system for a time $t$ up to 
$J_0 t_\final = (357~\mathrm{rad/s}) \times (15~\mathrm{ms}) 
\approx 5.4$~(Fig.~\ref{fig_setup}c).
The global system has largely equilibrated internally, and fluctuations are small, as shown in Sec.~\ref{sec_Dynamics}.
Finally, we measure the states of pairs of neighboring qubits via quantum state tomography: We measure the nontrivial two-qubit Pauli operators' expectation values across many trials~\cite[App.~G]{NYH_20_Noncommuting}.

\subsection{Initial state}
\label{sec_Init_State}

Conventional thermalization experiments begin with the global system in a microcanonical subspace, a joint eigenspace shared by the global charges (apart from the energy).
As our global charges do not commute,
they cannot have well-defined nonzero values simultaneously;
no nontrivial microcanonical subspace exists.
We therefore prepare the global system in an AMC subspace,
where the charges have fairly well-defined values.
We follow the proposal in~\cite{NYH_20_Noncommuting} for
extending the AMC subspace's definition, devised abstractly in~\cite{NYH_16_Microcanonical}, to realistic systems: 
In an AMC subspace, each global charge $Q_\gamma^\tot$
has a variance
$\sim O( \Sites^{\nu} )$,
wherein $\nu \leq 1$.
Every tensor product of single-qubit pure states
meets this requirement~\cite{NYH_20_Noncommuting}.

We choose the product to answer
an open question. In~\cite{NYH_20_Noncommuting},
$\rho_\NATS$ was found numerically to predict
a small system's long-time state best.
However, other thermal states approached $\rho_\NATS$ 
in accuracy as $\Sites$ grew.
(Accuracy was quantified with the long-time state's relative-entropy distance to a thermal state, as detailed in the next section.)
Does the NATS's accuracy remain greatest by an approximately constant amount,
as $\Sites$ grows, for any initial state?
The answer is yes for all $\Sites$ realized in our experiment.

The initial state,
\begin{align}
   \label{eq_Init_State}
   \ket{\psi_0}  
   := \ket{y+, x+, z+}^{\otimes \Sites / 3} \, ,
\end{align}
consistently distinguishes the NATS
for an intuitive reason synopsized here and detailed in App.~\ref{app_Initial_State}.
The initial state determines as follows the inverse temperature $\beta$ and chemical potentials $\mu_\gamma$ in  Eq.~\eqref{eq_NATS}~\cite{NYH_20_Noncommuting}.
Denote the global NATS by
$\rho_\NATS^\tot
:= \exp \left( - \beta \left[ H_\Heis 
   - \sum_{\gamma = x,y,z} \mu_\gamma S_\gamma^\tot
   \right] \right)
   / Z_\NATS^\tot$,
wherein $Z_\NATS^\tot$ normalizes the state.
$\beta$ and the $\mu_\gamma$'s are defined through~\cite{D'Alessio_16_From,NYH_20_Noncommuting}
\begin{align}
   \label{eq_beta_Def}
   & \bra{\psi_0} H_\Heis \ket{\psi_0}
   = \Tr \left( H_\Heis \, \rho_\NATS^\tot \right) 
   \quad \text{and} \\
   \label{eq_Mu_Def}
   & \bra{\psi_0} \sigma_\gamma^\tot \ket{\psi_0}
   = \Tr \left( \sigma_\gamma^\tot \, \rho_\NATS^\tot \right) 
   \quad \forall \gamma = x, y, z .
\end{align}
As the temperature approaches infinity, all thermal states converge to the maximally mixed state and so lose their distinguishability. We therefore choose the initial state such that $\beta$ is finite.
Additionally, the chemical potentials should be large, such that all noncommuting charges influence $\rho_\NATS$ substantially.
Upon choosing $\ket{\psi_0}$, we calculate $\beta$ and the $\mu$'s from Eqs.~\eqref{eq_Mu_Def} numerically by solving a maximum-entropy problem, following~\cite{Agmon_79_Algorithm, Alhassid_78_Upper}:
$\beta = 1.3 \times 10^{-3}$ \si{ s.rad^{-1}}, and
$\mu_{x,y,z} = -1046$ \si{ rad.s^{-1}}.

For generality, we have also tested other initial states: Permuting the factors in Eq.~\eqref{eq_Init_State}, we change the initial state's temperature. However, our qualitative conclusions continue to hold.


%
%
%
\section{Results}
\label{sec_Results}

Having introduced our setup and protocol, we observe, in Sec.~\ref{sec_Dynamics}, the dynamics of thermalization influenced by noncommuting charges. Section~\ref{sec_Vs_N} evidences thermalization to near $\rho_\NATS$.
Section~\ref{sec_Commut} compares these results with thermalization in the presence of just two commuting charges.

\subsection{Dynamics}
\label{sec_Dynamics}

\begin{figure}
\centering
    \includegraphics[width=88mm]{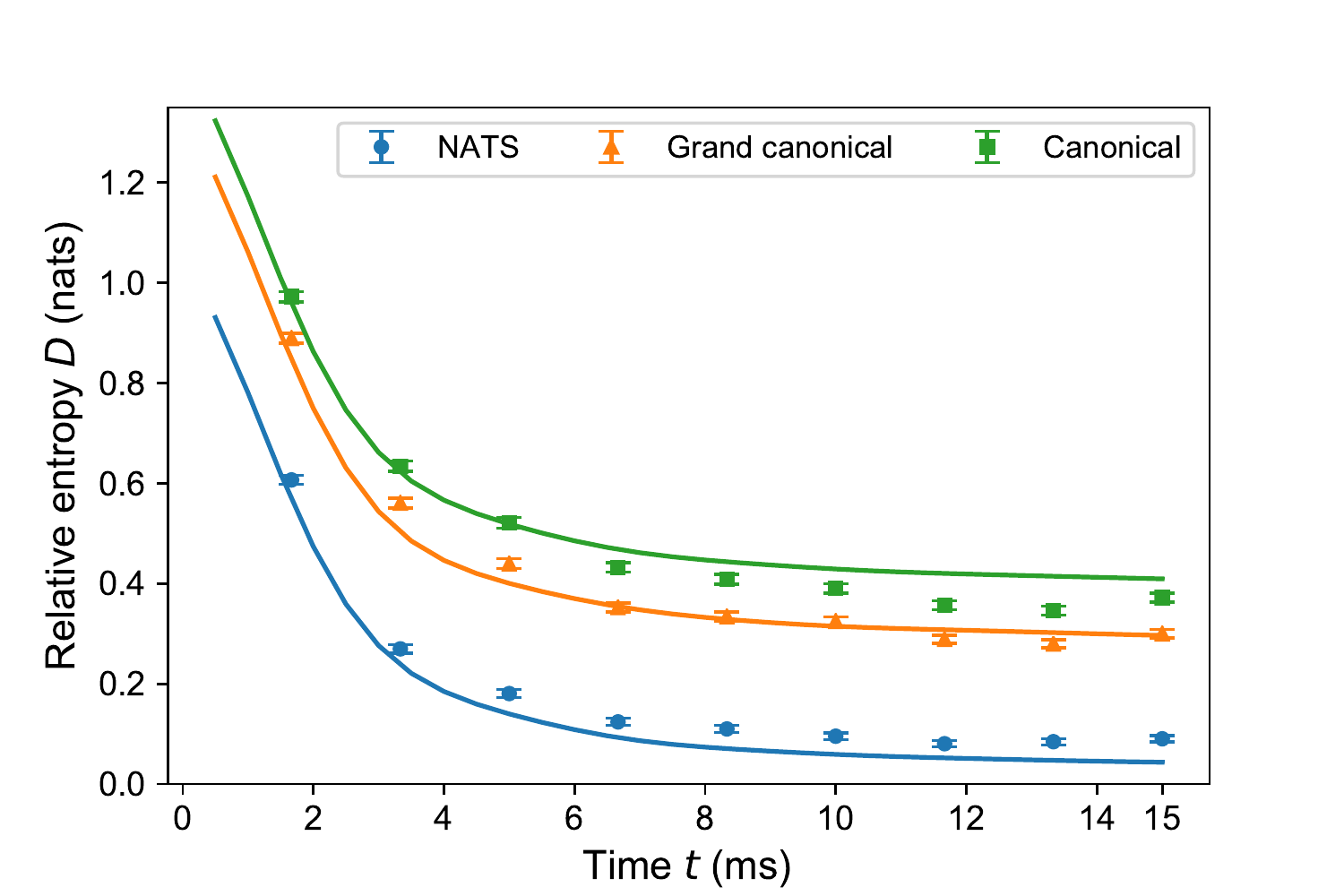}
    \caption{\caphead{Average distance from system-of-interest state to thermal prediction vs. time:} 
    The ion chain consists of $\Sites = 21$ qubits.
    Each nearest-neighbor pair forms a small system of interest.
    We measure its state's distance to the NATS (blue points), 
    using the relative entropy~\cite{NYH_16_Microcanonical}, 
    and average over the pairs in the chain.
    Markers show experimental data, while lines are calculated numerically from Eq.~\eqref{eq_Heis}.
    Each data point is formed from 250~repetitions.
    The error bars are estimated by bootstrapping~\cite{Efron1986}.
    Entropies are measured in units of nats (are base-$e$).
    We also compare the NATS prediction with two competitor thermal predictions, following~\cite{Rigol_07_Relaxation,Langen_15_Experimental,NYH_20_Noncommuting}: canonical and grand canonical states. 
    At all times, the NATS predicts the state unambiguously more accurately than the competitors do.
    }
    \label{fig_DRel_time}
\end{figure}

Figure~\ref{fig_DRel_time} shows how accurately the NATS predicts 
a small system's state, as a function of time.
The global system size is $\Sites = 21$.
To construct the blue dots,
we measure the time-dependent state $\rho_{t}^{(j,j+1)}$
of each nearest-neighbor qubit pair $(j, j+1)$,
for $j = 1, 2, \ldots, \Sites - 1$.
We then calculate the state's distance to the NATS, measured with the relative entropy used often in quantum information theory~\cite{NielsenC10}:
If $\chi$ and $\xi$ denote quantum states (density operators) defined on the same Hilbert space, the \emph{relative entropy} is
$D(\chi || \xi) = \Tr (\chi [\log \chi - \log \xi])$.
(All logarithms in this paper are base-$e$:
Entropies are measured in units of nats---not to be confused with the NATS---rather than in bits.)
The relative entropy boasts an operational interpretation:
$D(\chi || \xi)$ quantifies the optimal efficiency with which the states can be distinguished, on average,
in a binary hypothesis test~\cite{NielsenC10}.
The relative entropy to the NATS has been bounded with quantum-information-theoretic techniques~\cite{NYH_16_Microcanonical} and calculated numerically in simulations~\cite{NYH_20_Noncommuting}.
Appendix~\ref{app_Calc_Thermal_State} describes how we calculate $\rho_\NATS$ numerically. 
We average $D( \rho_{t}^{(j,j+1)} || \rho_\NATS )$
over the $\Sites - 1$ qubit pairs, producing
$\langle D( \rho_{t}^{(j,j+1)} || \rho_\NATS ) \rangle$.
To our knowledge, this is the first report on the process of quantum many-body thermalization colored by noncommuting charges (e.g., begun in an AMC subspace).

As in~\cite{Rigol_07_Relaxation,Langen_15_Experimental}, 
we compare the small system's state with
competing predictions by other thermal states:
the canonical state
$\rho_\can := e^{- \beta H} / Z_\can$
and the grand canonical state
$\rho_\GC := \exp \left(
- \beta \left\{ H - \mu_z S_z^{(2)}
\right\} \right) / Z_\GC$.
The partition functions $Z_\can$ and $Z_\GC$ normalize the states.
We have denoted by $H$ the two-site Hamiltonian and by $S_z^{(2)}$ the two-site spin operator.
We call $\rho_\GC$ ``grand canonical'' because $S_z$ is equivalent to a spinless-fermion particle-number operator via a Jordan--Wigner transformation~\cite{Jordan_28_Quantum}.
As the blue discs (distances to $\rho_\NATS$) are lower than the orange triangles ($\rho_\GC$) and green squares ($\rho_\can$),
the NATS always predicts the state best.

The curves show results from numerical simulations.
In the simulations, we exactly model time evolution under the Heisenberg Hamiltonian.
The experimental markers lie close to the theoretical curves.
Yet the distance to $\rho_\GC$ is slightly less empirically than theoretically, on average over time;
the same is true of $\rho_\can$;
and the opposite is true of $\rho_\NATS$.
These slight mismatches arise from noise, 
which we describe now.

As a real-world quantum system, 
the ion chain is open.
The environment affects the chain similarly to 
a depolarizing channel, 
which brings the state toward the maximally mixed state,
$\id / 2^\Sites$~\cite{NielsenC10}.
Of our candidate two-qubit thermal states,
$\rho_\can$ lies closest to 
$\id / 4$, 
$\rho_\GC$ lies second-closest,
and $\rho_\NATS$ lies farthest.
We can understand why information-theoretically~\cite{Jaynes_57_Information_II,Guryanova_16_Thermodynamics,Lostaglio_17_Thermodynamic}:
If one knows nothing about the system of interest,
one can mostly reasonably ascribe to the system
the state $\id / 4$.
Knowing nothing except the average energy, 
one should ascribe $\rho_\can$.
Knowing only the average energy and $\langle S_z^{(2)} \rangle$, one should ascribe $\rho_\GC$.
Knowing the average energy and $\langle S^{(2)}_{x,y,z} \rangle$, one should ascribe $\rho_\NATS$.
The more information a thermal state encodes,
the farther it is from $\id / 4$.
The depolarizing noise,
bringing the two ions' state closer to $\id / 4$,
brings the state closer to $\rho_\can$ and $\rho_\GC$
but not so close to $\rho_\NATS$
(in fact, away from $\rho_\NATS$, as explained in App.~\ref{app:DepolNoise}).
Hence the deviations between experimental markers and theoretical predictions in Fig.~\ref{fig_DRel_time}.

Nonetheless, the experiment exhibits considerable resilience to noise.
The chain can leak four charges ($S_{x,y,z}$ and energy) to its environment, violating the conservation laws ideally imposed on the ions. One might expect these many possible violations to prevent $\rho_\NATS$ from predicting the long-time state accurately. However, our results show otherwise: The chain is closed enough that $\rho_\NATS$, as a prediction, bests all competitor thermal states that may be reasonably expected from thermodynamics and information theory~\cite{Jaynes_57_Information_II}. Appendix~\ref{app:DepolNoise} supports this conclusion with simulations of depolarization atop the Trotterized Heisenberg evolution.

By $t_\final = 15$ ms, the curves in Fig.~\ref{fig_DRel_time} are approximately constant;
the small system has approximately thermalized. 
Thermalization occurs more completely at large $\Sites$ than at small $\Sites$, but 15 ms suffices for all the curves to drop substantially.
Our choice of experimental run time is thereby justified.
(For details about fluctuations in the relative entropy, see App.~\ref{app_Thermd}.)

\subsection{Thermalization to near the non-Abelian thermal state}
\label{sec_Vs_N}

\begin{figure}
\centering
    \includegraphics[width=88mm]{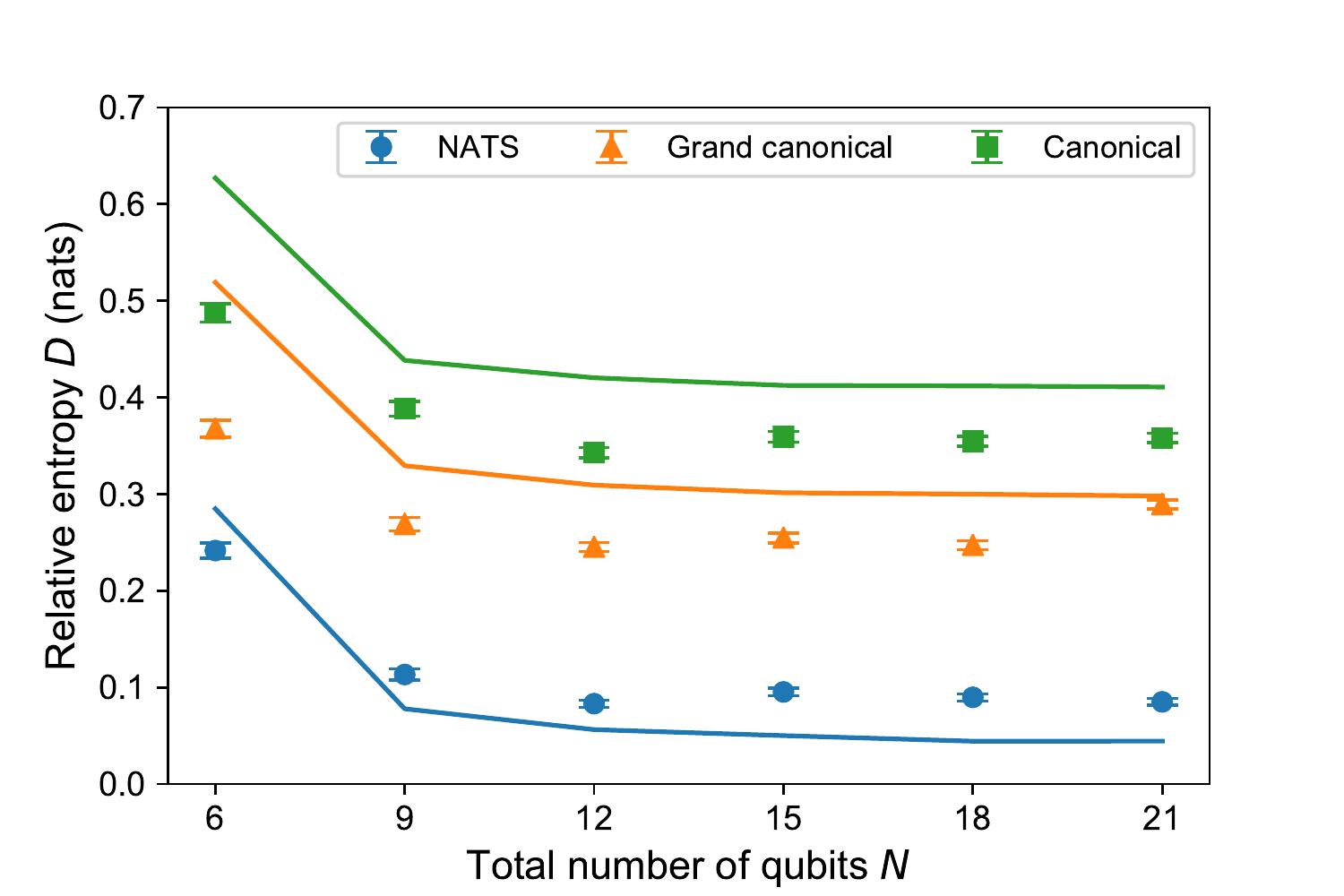}
    \caption{\caphead{Average distance from long-time system-of-interest state to thermal prediction vs. total number of qubits:}  Markers show experimental data, while lines are calculated numerically, using unitary dynamics, from Eq.~\eqref{eq_Heis}.
    The NATS predicts the final state the most accurately at all system sizes. Depolarizing noise appears to explain the experiment--theory discrepancies. The $\Sites=6$ point is an outlier due to the global system's small size.
    }
    \label{fig_Drel_to_exp_NrQubits}
\end{figure}

In Fig.~\ref{fig_Drel_to_exp_NrQubits}, 
we focus on late times 
while varying the global system size.
We average over the final three time points,
as the relative entropies have equilibrated 
but fluctuate slightly across that time (App.~\ref{app_Thermd}).
The blue discs represent the relative-entropy distance from 
the final system-of-interest state,
$\rho_{t_\final}^{(j,j+1)}$,
to the NATS, averaged over qubit pairs.
The average distance declines from 
0.24(2) nats to 0.085(6) nats as $\Sites$ grows from 6 to 21.
These values overestimate the true values by approximately 0.03 nats, because the number of experimental trials was finite.
For reference, $D(\chi || \xi)$ obeys no upper bound. We hence answer two open questions: Equilibration to near the NATS occurs in realistic systems and is experimentally observable, despite the opportunity for the spin chain to leak many charges via decoherence. Furthermore, the orange triangles (distances to $\rho_\GC$) lie 0.16 nats above the blue discs (distances to $\rho_\NATS$), on average; and the green squares (distances to $\rho_\can$) lie 0.26 nats above the blue discs, on average. Hence the NATS prediction is distinguishably most accurate at all experimentally realized $\Sites$.


Appendix~\ref{app_Cirac_Arg} analytically extends this conclusion beyond the experimental system sizes: Consider averaging each thermal state over the qubit pairs. The averaged $\rho_\NATS$ differs from the averaged competitor thermal states, as measured by nonzero relative-entropy distances. The distances are lower-bounded by a constant at all $\Sites$, even in the thermodynamic limit
(as $\Sites \to \infty$). We prove this claim about $\rho_\NATS$'s distinguishability under assumptions met by our experiment. 

We have observed equilibration to near the NATS, but the small system does not thermalize entirely:
$\langle D( \rho_{t_\final}^{(j,j+1)} || \rho_\NATS ) \rangle \neq 0$.
We expect the lingering athermality to stem partially from
the global system's finite size~\cite{Beugeling_14_Finite,Corps_21_Long}.
Yet charges' noncommutation has been conjectured 
to hinder thermalization additionally~\cite{NYH_20_Noncommuting}.
We now dig further into that conjecture.


\subsection{Comparison with commuting charges}
\label{sec_Commut}

Let us compare thermalization steered by noncommuting charges
with thermalization steered by just commuting charges. 
We realize the commuting case with the long-range $XY$ Hamiltonian
\begin{equation}
    \label{eq_XY_Hamiltonian}
    H_{xy} := \sum_{j<k} 
    \frac{1}{2} \frac{J_0}{\left| j-k \right|^\alpha} 
    \left( \sigma_x^{(j)} \sigma_x^{(k)} 
    + \sigma_y^{(j)} \sigma_y^{(k)} \right),
\end{equation}
for $\Sites = 21$, with $J_0 = 398$~rad/s
and $\alpha = 0.86$
(see Sec.~\ref{sec_Methods} for details). 
The charges are the total energy and $\sigma_z^\tot$.
We Trotter-approximate $H_{xy}$ similarly to $H_\Heis$
(App.~\ref{app_Trotter}).
We prepare 
$\ket{y+, x+, z+}^{\otimes \Sites / 3}$,
such that the commuting-charge experiment parallels the noncommuting-charge experiment (which begins in an AMC subspace, too) as closely as possible. Then,
we simulate $H_{xy}$ for 10~ms.\footnote{
We chose the evolution times such that the system effectively evolved under the $XY$ and Heisenberg models for the same amount of time. We simulated the Heisenberg model by repeating three Trotter steps ($XX$, $YY$ and $ZZ$). Therefore, during a Trotter-sequence evolution of 15 ms, the system effectively evolved under a Heisenberg model for 5 ms. Simulating the $XY$ model, we repeated only two Trotter steps ($XX$ and $YY$), requiring a total time of $2 \times 5$ ms $= 10$ ms.}

\begin{table}[]
\begin{tabular}{lllllll}
\hline
\multicolumn{1}{|l|}{N} &
  \multicolumn{1}{l|}{6} &
  \multicolumn{1}{l|}{9} &
  \multicolumn{1}{l|}{12} &
  \multicolumn{1}{l|}{15} &
  \multicolumn{1}{l|}{18} &
  \multicolumn{1}{l|}{21} \\ \hline
  \multicolumn{1}{|l|}{D} &
  \multicolumn{1}{l|}{\begin{tabular}[c]{@{}l@{}}0.19(2)\end{tabular}} &
  \multicolumn{1}{l|}{\begin{tabular}[c]{@{}l@{}}0.096(10)\end{tabular}} &
  \multicolumn{1}{l|}{\begin{tabular}[c]{@{}l@{}}0.077(6)\end{tabular}} &
  \multicolumn{1}{l|}{\begin{tabular}[c]{@{}l@{}}0.066(7)\end{tabular}} &
  \multicolumn{1}{l|}{\begin{tabular}[c]{@{}l@{}}0.057(5)\end{tabular}} &
  \multicolumn{1}{l|}{\begin{tabular}[c]{@{}l@{}}0.056(6)\end{tabular}} \\ \hline
 &
   &
   &
   &
   &
   &
   \\
   &   &   &   &   &   &
\end{tabular}
    \caption{\caphead{Average distance from long-time system-of-interest state to grand canonical state, $\rho_\GC$, when only energy and $\sigma_z^\tot$ conserved:} 
    Each nearest-neighbor pair in the ion chain forms 
    a small system of interest. The pair's long-time state lies some distance from $\rho_\GC$ (the most accurate thermal prediction). We measure this distance using the relative entropy $D$~\cite{NYH_16_Microcanonical}, 
    measured in nats. Then, we average over the pairs in the chain.
    The average pair thermalizes more thoroughly than if noncommuting charges are conserved, at all global system sizes~$\Sites$.} 
    \label{tbl_Drel_to_exp_NrQubits_XY}
\end{table}

Table \ref{tbl_Drel_to_exp_NrQubits_XY} shows the results. 
The average small system thermalizes more thoroughly when determined by commuting charges than when determined by noncommuting charges. For instance, in the commuting case, 
the relative entropy to $\rho_\GC$ descends as low as 0.056(6) nats,when $\Sites = 21$. 
In the noncommuting case, when $\Sites = 21$, the relative entropy to $\rho_\NATS$ reaches $0.085(6) > 0.056(6)$ nats. 
This result is consistent with the conjecture that charges' noncommutation hinders thermalization~\cite{NYH_20_Noncommuting}, as well as with the expectation that, in finite-size global systems, a small system's long-time entanglement entropy decreases as the number of charges grows~\cite{Monteiro_21_Quantum}.
Future work will distinguish how much our charges' noncommutation is hindering thermalization and how much the multiplicity of charges is.

\section{Conclusions}
\label{sec_Conclusions}

We have observed the first experimental evidence of
a particularly quantum equilibrium state:
the non-Abelian thermal state, 
which depends on noncommuting charges. 
Whereas typical many-body experiments begin
in a microcanonical subspace,
our experiment begins in an approximate microcanonical subspace.
This generalization accommodates 
the noncommuting charges' inability to have 
well-defined nontrivial values simultaneously.
Our experiment affirmatively answers an open question: 
whether, for any initial state, 
the NATS remains a substantially better prediction than 
other thermal states as the global system grows.
Our trapped-ion experiment affirmatively answers 
two more open questions:
(i) whether realistic systems exhibit 
the thermodynamics of noncommuting charges
and (ii) whether this thermodynamics can be observed experimentally, despite the abundance of the conservation laws that decoherence can break.
Our work therefore bridges quantum simulators to 
the emerging subfield of noncommuting charges 
in quantum-information thermodynamics.
The subfield has remained theoretical until now; 
hence many predictions now can, and should, be tested experimentally---predictions about reference frames, second laws of thermodynamics, information storage in dynamical fixed points, and more~\cite{Lostaglio_17_Thermodynamic, Guryanova_16_Thermodynamics, NYH_18_Beyond, Sparaciari_18_First, Khanian_20_From, Khanian_20_Resource, Gour_18_Quantum, Manzano_22_Non, Popescu_18_Quantum, Popescu_19_Reference, Lostaglio_14_Masters, NYH_16_Microcanonical, Ito_18_Optimal, Bera_19_Thermo, Mur_Petit_18_Revealing, Manzano_18_Squeezed,  NYH_20_Noncommuting, Manzano_20_Hybrid, Fukai_20_Noncommutative, Mur_Petit_19_Fluctuations, Scandi_18_Thermodynamic, Manzano_18_Squeezed, Sparaciari_18_First, Mur_Petit_18_Revealing, Boes_18_Statistical, Ito_18_Optimal, Mitsuhashi_22_Characterizing, Croucher_18_Information, Vaccaro_11_Information, Wright_18_Quantum,Zhang_20_Stationary,Zhang_20_Stationary, Medenjak_20_Isolated,Croucher_21_Memory,NYH_22_How,Marvian_21_Qudit,Marvian_22_Rotationally,Ducuara_22_Quantum,Murthy_22_Non,Majidy_22_Non}.

In addition to answering open questions, our results open avenues for future work.
First, Fig.~\ref{fig_Drel_to_exp_NrQubits} contains blue discs (distances to $\rho_\NATS$) that could be fitted. The best-fit line could be compared with the numerical prediction in~\cite{NYH_16_Microcanonical} and the information-theoretic bound in~\cite{NYH_20_Noncommuting}. Obtaining a reliable fit would require the reduction of systematic errors, such as decoherence, and the performance of more trials. 
Second, we observed that the small system thermalizes less in the presence of noncommuting charges than in the presence of just commuting charges.  
Future study will tease apart effects of the charges' noncommutation from effects of the charges' multiplicity.

Third, the quantum-simulation toolkit developed here
merits application to other experiments.
We combined our quantum simulator's native interaction
with rotations and dynamical decoupling
to simulate a non-native Heisenberg interaction.
The Trotterized long-range Hamiltonian, with the single-qubit control used to initialise our system, can be advantageous for studying more many-body physics with quantum simulators.
As our experiment reached system sizes larger than can reasonably be simulated realistically (including noise), our toolkit's usefulness in many-body physics is evident.
These techniques can be leveraged to explore
nonequilibrium Heisenberg dynamics~\cite{Geier2021Floquet,Scholl:2022}, 
topological excitations~\cite{birnkammer2022}, and more.
Beyond the Heisenberg model, the impact of charges' noncommutation on equilibration can be studied in more-exotic contexts, such as lattice gauge theories~\cite{Mueller_22_Thermalization,Zhou_22_Thermalization}.

%
%
\begin{acknowledgments}
N.Y.H. thanks Michael~Beverland, Ignacio~Cirac,
Markus~Greiner, Julian~L\'{e}onard, Mikhail~Lukin, Vladan~Vuletic, and Peter~Zoller for insightful conversations.
N.Y.H. and M.J. thank Daniel James, Aephraim Steinberg, and their compatriots for co-organizing the 2019 Fields Institute conference at which this collaboration began.
A.L. thanks Christopher D. White for suggestions about numerical techniques.
The project leading to this application received funding from the European Union’s Horizon 2020 research and innovation programme under grant agreement No 817482. Furthermore, we acknowledge support from the Austrian Science Fund through the SFB BeyondC (F7110), funding by the Institut f\"ur Quanteninformation GmbH, and support from the John Templeton Foundation (award no. 62422). The opinions expressed in this publication are those of the authors and do not necessarily reflect the views of the John Templeton Foundation or UMD.
This research was supported by the National Science Foundation under Grant No. NSF PHY-1748958; 
Grant No. NSF PHY-1748958; 
and an NSF grant for the Institute for Theoretical Atomic, Molecular, and Optical Physics at Harvard University and the Smithsonian Astrophysical Observatory.
\end{acknowledgments}

\begin{appendices}
%
\renewcommand{\thesection}{\Alph{section}}
\renewcommand{\thesubsection}{\Alph{section} \arabic{subsection}}
\renewcommand{\thesubsubsection}{\Alph{section} \arabic{subsection} \roman{subsubsection}}

\makeatletter\@addtoreset{equation}{section}
\def\theequation{\thesection\arabic{equation}}

\section{Methods}
\label{sec_Methods}

This section provides details about the setup (Sec.~\ref{sec_Setup_Methods}), the realization of spin--spin interactions (Sec.~\ref{sec_Interactions_Methods}), the Trotterization of the Heisenberg Hamiltonian (Sec.~\ref{sec_Trotter_Methods}), and the quantum state tomography and statistical analysis (Sec.~\ref{sec_Tomog_Methods}).

\subsection{Experimental setup}
\label{sec_Setup_Methods}

A linear ion crystal of 21 ${}^{40}\mathrm{Ca}^+$ ions is trapped in a linear Paul trap with trapping frequencies of $\omega_x=2\pi \times 2.930$~MHz (radially) and $\omega_\mathrm{ax}=2\pi \times 0.217$~MHz (axially). The qubit states $\ket{z+}$ and $\ket{z-}$ are coupled by an optical quadrupole transition, which we drive with a titanium-sapphire laser, with a sub--10~Hz linewidth, at 729~nm. Collective qubit operations are implemented with a resonant beam that couples to all the qubits with approximately equal strengths. Single-qubit operations are performed with a steerable, tightly focused beam that induces AC Stark shifts. 
In some trials, the system size $\Sites$ is less than 21. In these cases, we hide the unused ions in the Zeeman sublevel $\left| {3}^{2}\mathrm{D}_{5/2}, m=-3/2 \right>$.

Recall that the initial state is ideally the product $\ket{\psi_0}$ in Eq.~\eqref{eq_Init_State}.
The experimental initial state $\ket{ \psi_\mathrm{exp} }$ has a fidelity
$|\left< \psi_\mathrm{exp} | \psi_0 \right>|^2=0.90(2)$
for $N=21$. In each experimental cycle, we cool the ions via Doppler cooling and polarisation-gradient cooling~\cite{Joshi2020PGC}. We also sideband-cool all transverse collective motional modes to near their ground states. Then, we prepare the state~\eqref{eq_Init_State}, simulate the Heisenberg evolution, and measure the state. The cycle is repeated 300--500 times per quantum-state-tomography measurement basis.

\subsection{Implementing the effective Heisenberg interaction}
\label{sec_Interactions_Methods}

We implement the long-range spin--spin interaction~\eqref{eq_IsingH} with a  laser beam carrying two frequencies that couple motional and electronic degrees of freedom of the ion chain. The beam's frequency components, $\omega_\pm = \pm(\omega_x+\Delta)$, are symmetrically detuned by $\Delta=2\pi \times 27$~kHz (for $\Sites=21$ ions) from the transverse--center-of-mass mode, which has a frequency
$\omega_x=2\pi \times 2.930$~MHz.
A third frequency component, $\omega_\mathrm{AC}=2\pi\times1.4~\mathrm{MHz}$,  is added to the bichromat beam. This component compensates for the additional AC-Stark shift caused by other electronic states~\cite{jurcevic2014quasiparticle}. 

The resulting spin--spin coupling effects a long-range Ising model,
$\sum_{j<k} J_{j,k} 
    \sigma_x^{(j)} \sigma_x^{(k)} \, .$
The $J_{j,k}$ denotes the strength of the coupling between ions $j$ and $k$.
$J_{j,k}$ approximates the power law in Eq.~\eqref{eq_IsingH}, where the coupling strength equals $J_0=468$~rad/s and the exponent $\alpha=0.86$ for the 21-ion chain. 

Directly realizing the desired long-range Heisenberg Hamiltonian~\eqref{eq_Heis} for trapped ions is difficult~\cite{Porras:2004,Grass:2014}. 
Instead, we simulate $H_\Heis$ via Trotterization. 
After the first time step, we change the interaction from $H_{xx}$ to $H_{yy}$; after the second time step, to $H_{zz}$; and, after the third time step, back to $H_{xx}$. We perform this cycle, or \emph{Trotter step}, $\numSteps$ times~\cite{lloyd1996}. We can realize $H_{yy}$ by shifting the bichromat light's phase by $\pi/2$ relative to the phase used to realize $H_{xx}$. Implementing $H_{zz}$ requires a global rotation: Denote by $R_y$ a $\pi/2$ rotation of all the qubits about the $y$-axis. We can effect $H_{zz}$ with, e.g., 
$R_y^\dag H_{xx} R_y$.

\subsection{Noise-robust Trotter sequence}
\label{sec_Trotter_Methods}

In our experimental setup, most native decoherence is dephasing relative to the $\sigma_z$ eigenbasis, which rotations transform into effective depolarization (App.~\ref{app:DepolNoise}). 
This noise results from temporal fluctuations of (i) the magnetic field and (ii) the frequency of the laser that drives the qubits.
Earlier experiments on this platform involved $XY$-interactions, which enable the quantum state to stay in a decoherence-free subspace~\cite{jurcevic2014quasiparticle,Brydges_19_Probing,Joshi2021}. 
Here, the dynamics must be shielded from dephasing differently. 
We mitigate magnetic-field noise by incorporating a dynamical-decoupling scheme into the Trotter sequence (Fig.~\ref{fig_setup}b). Furthermore, we design the Trotter sequence to minimise the number of global rotations. This minimisation suppresses the error accumulated across all the rotations. We reduce this error further by alternating the rotations' directions between Trotter steps. For further details, see App.~\ref{app_Trotter}.

To formalise the Trotter sequence, we introduce notation:
Let $U_{xx}=\exp(-i H_{xx} t)$ and $U_{yy}=\exp(-i H_{yy}t)$.
For $\gamma \in \{x, y\}$, $R_\gamma
:=  \exp \left(- i \, \frac{\pi}{4} \, \sigma_\gamma^\tot  \right)$
denotes a global $\pi/2$ rotation about the $\gamma$-axis. $\numSteps$ denotes the number of Trotter steps. Each Trotter step consists of either the operation
$E_+ = U_{yy} U_{xx} R_x U_{yy}$ or the operation
$E_- = U_{yy} U_{xx} R_x^\dag U_{yy}$.
To simulate a Heisenberg evolution for a time $t$,
we implement the Trotter sequence
\begin{align}
   U_\Heis (t) 
   \approx  R_y^\dag  R_x
   \left[ (E_-)^4  (E_+)^4  \right]^{\numSteps / 8}
   R_x^\dag  R_y .
\end{align}
This sequence protects against decoherence and over-/under-rotation errors caused by global pulses. Numerical simulations supporting this claim appear in App.~\ref{app:NumSimDD}.    

The Trotter sequence lasts for 15~ms, containing $\leq 36$ Trotter steps. Each Trotter step consists of three substeps, each lasting for approximately 139~$\mu$s. Each substep's rising and falling slopes are pulse-shaped to avoid incoherent excitations of vibrational sidebands of the qubit transition. The slopes reduce the effective spin--spin coupling by a factor of 0.84, and the actual interaction time is 115~$\mu$s. Thus, the effective spin--spin coupling values used in Eq.~\eqref{eq_Heis} ranged from $J_0 = 336~\mathrm{rad/s}$ for 6~qubits and $J_0 = 398~\mathrm{rad/s}$ for 21~qubits.

The magnetic-field variations occur predominantly at temporally stable 50-Hz harmonics. We reduce the resulting Zeeman-level shifts via feed-forward to a field-compensation coil \cite{kranzl2022}. The amplitudes end up below 3~Hz, for all 50-Hz harmonics between 50~Hz and 900~Hz. 
Consider a simple Ramsey experiment on the qubit transition $4^2\mathrm{S}_{1/2} (m=+1/2) \leftrightarrow 3^2\mathrm{D}_{5/2} (m=+5/2)$. The corresponding $(1/e)$-contrast coherence time is 47(6)~ms. The global qubit rotations are driven by the elliptically shaped 729-nm beam, which causes spatially 	
inhomogeneous Rabi frequencies that vary across the ion crystal by 6\%.

\subsection{Quantum state tomography}
\label{sec_Tomog_Methods}

We measure each qubit pair's state via quantum state tomography. In each measurement basis, 300--500 quantum state measurements were carried out. To reconstruct the state from the measurements, we used maximum-likelihood estimation~\cite{Jezek2003}. We estimated statistical uncertainties by bootstrapping~\cite{Efron1986}.

\section{Rate of hopping during Heisenberg evolution} 
\label{app_Hopping}

In this section, we derive an expression for the Heisenberg Hamiltonian's spin-exchange rate. For simplicity, we model two qubits governed by the Heisenberg Hamiltonian
\begin{equation}
    H=\frac{J_0}{3} \left(
    \sigma_x^{(1)}\sigma_x^{(2)}
    +\sigma_y^{(1)}\sigma_y^{(2)}
    +\sigma_z^{(1)}\sigma_z^{(2)} \right).
\end{equation}
We relabel the $\sigma_z$ eigenstates as 
$\ket{z+} \equiv \ket{\uparrow}$ and 
$\ket{z-} \equiv \ket{\downarrow}$.
Matrices are expressed relative to the basis
formed from products of $\ket{\uparrow}$ and $\ket{\downarrow}$.
The Hamiltonian can be expressed as 
\begin{equation}
    H= \frac{J_0}{3} 
    \begin{pmatrix}
1 & 0 & 0  & 0\\
0 & -1 & 2 & 0\\
0 & 2 & -1 & 0\\
0 & 0 & 0 & 1
\end{pmatrix} ;
\end{equation}
and a pure two-qubit state, as
$\ket{\psi(t)} 
   = c_1(t) \ket{\downarrow \downarrow}
   + c_2(t) \ket{\uparrow \downarrow}
   + c_3(t) \ket{ \downarrow \uparrow} 
   +c_4(t) \ket{ \uparrow \uparrow} .$
The coefficients $c_k(t) \in \mathbb{C}$ depend on the time, $t$, and are normalized as $\sum_{k = 1}^4 | c_k(t) |^2 = 1$. 
The dynamics obey the Schr{\"o}dinger equation,
$H\ket{\psi(t)} 
= i\hbar \frac{d \ket{\psi(t)}}{dt} \, .$
Defining $\Omega := \frac{J_0}{3} $ and setting $\hbar = 1$,
we express the Schr{\"o}dinger equation in matrix form as
\begin{equation}
    \begin{pmatrix}
\dot{c}_1(t)\\
\dot{c}_2(t)\\
\dot{c}_3(t)\\
\dot{c}_4(t)
\end{pmatrix}
    = -i\Omega
    \begin{pmatrix}
1 & 0 & 0  & 0\\
0 & -1 & 2 & 0\\
0 & 2 & -1 & 0\\
0 & 0 & 0 & 1
\end{pmatrix}
\begin{pmatrix}
c_1(t)\\
c_2(t)\\
c_3(t)\\
c_4(t)
\end{pmatrix}.
\end{equation}
The solution is
\begin{eqnarray}
   \begin{cases}
   c_1(t) = c_1(0) e^{-i\Omega t} \\
   c_2(t) = \frac{1}{2} c_2(0)e^{-i\Omega t}(1+e^{4i\Omega t}) +\frac{1}{2} c_3(0) e^{-i\Omega t}(1-e^{4i\Omega t}) \\
   c_3(t) = \frac{1}{2} c_3(0)e^{-i\Omega t}(1+e^{4i\Omega t}) +\frac{1}{2} c_2(0) e^{-i\Omega t}(1-e^{4i\Omega t}) \\
   c_4(t) = c_4(0) e^{-i \Omega t}
   \end{cases} .
\end{eqnarray}

We aim to derive the time required for 
$\ket{\uparrow \downarrow}$ to transform into
$\ket{\downarrow \uparrow}$.
If the initial state is
$\ket{\psi(0)} = \ket{\uparrow \downarrow}$, 
the solution reduces to 
\begin{eqnarray}
   \begin{cases}
   c_1(t) = 0 \\
   c_2(t) = \frac{1}{2}e^{-i\Omega t} (1+e^{4i\Omega t})
   \nonumber \\
   c_3(t) = \frac{1}{2}e^{-i\Omega t} (1-e^{4i\Omega t}) \\
   c_4(t) = 0
   \end{cases} .
\end{eqnarray}
Consider measuring the $\sigma_z$ product eigenbasis at time $t$.
The possible outcomes $\uparrow \downarrow$ and $\downarrow \uparrow$ result with probabilities
\begin{align}
   & |c_2(t)|^2 
   = \frac{1}{4}(2+e^{4 i \Omega t} + e^{-4 i \Omega t}) 
   = \frac{1}{2} + \frac{1}{2} \cos(4 \Omega t)
   \quad \text{and} \\
   & |c_3(t)|^2 =  1-|c_2(t)|^2. 
\end{align}
Therefore, the two-qubit excitation-hopping frequency is $\Gamma_{\rm flip-flop} = 4\Omega  = 4 J_0/3 $. 
The corresponding period is defined as the hopping time:
\begin{equation}
    T_{\rm hop} = \frac{2\pi}{ 2\Gamma_{\rm flip-flop} } 
    = \frac{3\pi}{ 4J_0}. 
\end{equation}
This result agrees with our experimental results and 
can be extended simply to $\Sites$ qubits. 
If the interaction is nearest-neighbor only,
the time for hopping from site 1 to site $\Sites$ is
$T_{\rm hop} = (\Sites - 1) \frac{3\pi}{ 4J_0}$. 

\section{Initial state}
\label{app_Initial_State}

If the global system is prepared in $\ket{\psi_0}$ [Eq.~\eqref{eq_Init_State}], 
$\rho_\NATS$ models a small system's long-time state
distinctly more accurately than other thermal states
($\rho_\can$ and $\rho_\GC$) do,
at all the global system sizes $\Sites$ realized
(Fig.~\ref{fig_Drel_to_exp_NrQubits}).
Equation~\eqref{eq_Init_State}
distinguishes $\rho_\NATS$ for two reasons.

First, suppose that the temperature is high ($\beta \gtrsim 0$).
All the thermal states resemble 
the maximally mixed (infinite-temperature) state
$\id / 2^\Sites$
and so resemble each other.
We therefore keep the temperature low,
by keeping each charge's spatial density low:
We separate the $\ket{\gamma+}$'s from each other maximally, 
for each $\gamma = x,y,z$.
To provide a sense of $\beta$'s size at $\Sites = 12$,
we compare with the bandwidth of the Heisenberg Hamiltonian~\eqref{eq_Heis}, the greatest energy minus the least.
$\beta$ equals $7.13$ times the bandwidth's inverse 
and $1.74 \times 10^{-3} $ times the average energy gap's inverse.

Second, noncommuting charges distinguish NATS thermodynamics
from more-classical thermodynamics.
If we are to observe NATS physics, therefore, 
$\rho_\NATS$ [Eq.~\eqref{eq_NATS}] should depend significantly on $Q_\gamma^\tot$, for all $\gamma$.
Hence the $\mu_\gamma$'s should have large magnitudes---and 
so should the expectation values
$\bra{\psi_0} \sigma_\gamma^\tot \ket{\psi_0}$,
by Eq.~\eqref{eq_Mu_Def}.
Hence, for each $\gamma$, 
the $\sigma_\gamma$ eigenstates in $\ket{\psi_0}$
should be identical.
The ordering of the $x+$, $y+$, and $z+$
in Eq.~\eqref{eq_Init_State} does not matter.
Importantly, $\ket{\psi_0}$ is not 
an eigenstate of any $Q_\gamma^\tot$;
so the global system does not begin in a microcanonical subspace; so the experiment is not equivalent, by any global rotation,
to any experiment that conserves just $\sigma_z^\tot$ and that leads to $\rho_\GC$.
When $\Sites = 12$, $\beta \mu_z$ equals $-1.36 $ times
the inverse of each nonzero gap of $S_z^\tot$.

We numerically identified many tensor products of $\ket{\gamma \pm}$'s, as well as superpositions of energy eigenstates, that have $\beta$'s much greater than our $\beta$. These states suffer from drawbacks that render the states unsuitable for observing the NATS: Either $\mu_{x,y,z} = 0$ or only one of the three charges has a nonzero expectation value. Such states provide little direction information about noncommutating charges. Furthermore, the states are highly entangled and so are difficult to prepare experimentally. $\ket{\psi_0}$ is easy to generate, aside from having a large $\sum_\gamma \beta \mu_\gamma S_\gamma^{\mathrm{tot}}$.

\section{Numerical calculation of the non-Abelian thermal state}
\label{app_Calc_Thermal_State}

Consider calculating the NATS for qubits $j$ and $j+1$. One might substitute two-qubit observables into Eq.~\eqref{eq_NATS}. This substituting yields an accurate prediction in the weak-coupling limit~\cite{NYH_20_Noncommuting} However, the experiment's long-range interactions render a many-body-physics approach more accurate~\cite{D'Alessio_16_From}.
We calculate $\beta$ and $\mu_{x,y,z}$ from the definitions~\eqref{eq_beta_Def} and~\eqref{eq_Mu_Def},
which depend on whole-system observables.
Then, we construct the whole-system NATS in those equations, 
$\rho_\NATS^\tot
:= \exp \left( - \beta \left[ H_\Heis 
   - \sum_{\gamma = x,y,z} \mu_\gamma S_\gamma^\tot
   \right] \right)
   / Z_\NATS^\tot$.
Finally, we trace out all the qubits except for $j$ and $j+1$. 

We perform the trace stochastically~\cite{hutchpp},
for computational feasibility.
The stochastic trace requires an average over states selected Haar-randomly from the traced-out subspace.
We averaged over 50--1,000 samples,
the precise number determined for each $\Sites$ as follows.
First, for small $\Sites$, 
we calculated the trace exactly.
We then determined the number of samples required for our stochastic approximation to converge to the exact value. From this number of samples, we estimated the number required for greater $\Sites$. (To estimate, we scaled down the sample size approximately inversely proportionally with the dimensionality of the traced-out Hilbert space, erring on the side of using more samples than necessary.)
We sampled this many Haar-random states and approximated the trace stochastically. Then, we slightly increased the number of samples, approximated the trace stochastically again, and confirmed that the result did not change significantly.

\section{Effect of depolarizing noise on relative entropy}
\label{app:DepolNoise}

We expect depolarization to dominate our experiment's noise. The reason is the experimental Hamiltonian and Trotter sequence,  described in App. \ref{app_Trotter_Derive}, as well as the dominant native decoherence. We rotated the qubits to effectively transform the native $\sigma_x \sigma_x$ coupling into the Heisenberg Hamiltonian, which is isotropic. Meanwhile, dephasing relative to the $\sigma_z$ eigenbasis dominated the native decoherence. The rotations spread the dephasing errors to the $x$-, $y$-, and $z$-directions uniformly. Such isotropic errors effect depolarization~\cite{NielsenC10}.
This appendix reports on numerical simulations of depolarized Trotter evolutions. We infer that noise should not significantly affect the conclusions drawn from our experimental observations.

We simulated the Trotterized Heisenberg-Hamiltonian evolution, with and without depolarization, of 12 qubits. Depolarization probabilistically interchanges the 12-qubit state $\rho$ with the maximally mixed state:
$\rho \mapsto \mathcal{E}(\rho) 
= (1-p) \rho + p \frac{\id}{4}$.
We chose for the noise parameter $p$ to equal 0.06, and we applied the channel $\mathcal{E}$ every 1.5 ms. 
This $p$ value is 30 times higher than the value that best reproduces the experimental state's distance from $\rho_\GC$.
We simulated an evolution of 45 ms. 

\begin{figure}
    \centering
    \includegraphics[width=88mm]{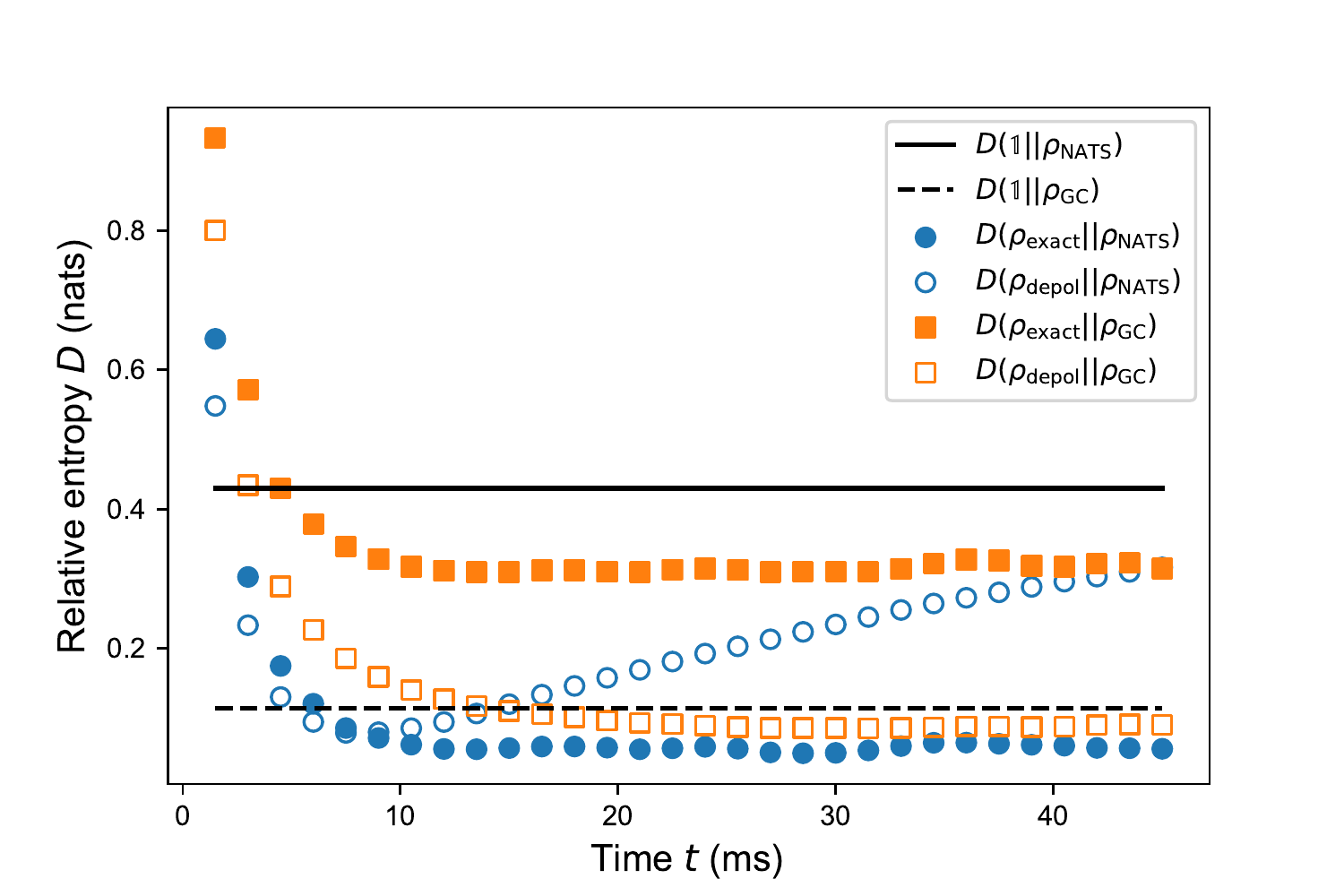}
    \caption{\caphead{Relative entropies from simulations with and without depolarizing noise vs. time:} 
    We simulated 12 qubits subject to Trotterized Heisenberg evolution alone (filled markers) or with depolarization (empty markers). Blue circles show relative-entropy distances to the NATS, and orange squares show distances to the grand canonical state.}
    \label{fig_Depol_noise}
\end{figure}

Figure~\ref{fig_Depol_noise} depicts the simulation's results. Time runs along the $x$-axis. Along the $y$-axis is the relative entropy between a system-of-interest state and a thermal state, averaged over all the chain's nearest-neighbor qubit pairs
(Sec.~\ref{sec_Dynamics}). 
$\rho_\exact$ denotes the final state of the depolarization-free simulation, and $\rho_\depol$ denotes the final state of the noisy simulation. We refer to the two states collectively as $\rho$.
We plot each state's distance to the NATS and distance to the grand canonical state.
We omit $\rho_\can$ for conciseness, although we analyzed this state, too. All qualitative conclusions about $\rho_\GC$ apply to $\rho_\can$.
The simulation is intended to reproduce qualitative effects, rather than exact experimental numbers, as we lack independent quantitative evaluations of the experimental noise. 

First, depolarizing noise affects 
$D( \rho || \rho_\NATS)$ oppositely 
$D( \rho || \rho_\GC)$. The reason is, depolarization transforms the simulated state into the maximally mixed state. $\rho_\GC$ lies close to $\id / 4$, closer than $\rho_\exact$ lies to $\rho_\GC$: In Fig.~\ref{fig_Depol_noise}, the dashed black line lies below the solid square markers at most times. (Section~\ref{sec_Dynamics} explains why.) Therefore, pushing $\rho$ toward $\id/4$,
depolarization pushes $\rho$ farther toward $\rho_\GC$ (nudges the empty square markers downward from the solid square markers, toward the dashed black line). In contrast, $\rho_\NATS$ lies farther from $\id / 4$ than $\rho_\exact$ lies from $\rho_\NATS$: The solid black line lies above the filled discs at most times. (Again, Sec.~\ref{sec_Dynamics} explains why.) Therefore, pushing $\rho$ toward $\id/4$, depolarization pushes $\rho$ farther from $\rho_\NATS$ (nudges the empty circles upward from the filled discs, toward the solid black line). Hence depolarization increases $D( \rho || \rho_\NATS)$ while decreasing 
$D( \rho || \rho_\GC)$.

Second, depolarization appears to affect 
$D( \rho || \rho_\NATS)$ more slowly than it affects 
$D( \rho || \rho_\GC)$. The reason is, depolarization pushes $D( \rho || \rho_\GC)$ in the same direction as the Trotterized Heisenberg evolution---downward. Therefore, $D( \rho || \rho_\GC)$ decreases quickly. In contrast, depolarization competes with the Heisenberg evolution in pushing $D( \rho || \rho_\NATS)$ downward. This competition makes $D( \rho_\depol || \rho_\NATS)$ depart from 
$D( \rho_\exact || \rho_\NATS)$ slowly; 
the unfilled circles in Fig.~\ref{fig_Depol_noise} separate from the filled discs more slowly than the unfilled square markers separate from the filled square markers.

Third, although depolarization ultimately raises 
$D( \rho_\depol || \rho_\NATS)$
well above $D( \rho_\exact || \rho_\NATS)$
in our simulation,
no such dramatic raising is visible in the experimental plot Fig.~\ref{fig_DRel_time}.
That is, the empty circles in Fig.~\ref{fig_Depol_noise} rise well above the filled discs; yet the blue discs in Fig.~\ref{fig_DRel_time} scarcely rise at the end of the experiment. Therefore, the experimental noise is weak and does not substantially affect 
$\langle D( \rho_t^{(j,j+1)} || \rho_\NATS) \rangle$.

Overall, the noise simulation affirms our main conclusion. We expect the experimental noise not to affect 
$D(\rho || \rho_\NATS)$ significantly while lowering 
$D(\rho || \rho_\GC)$ somewhat. Regardless of noise, the simulated state $\rho_{\exact / \depol}$ lies closest to $\rho_\NATS$ by a large margin. We can therefore have confidence that the NATS's predictive accuracy does not stem from the dominant noise.

Simulating the $XY$ model yields different results but the same conclusion: Noise affects the experimental results insignificantly. The XY model conserves only two charges ($\sigma_z^\tot$ and the Hamiltonian), so $\rho_\NATS$ should not predict the long-time state accurately. Indeed, 
$D(\rho || \rho_\NATS ) > D(\rho || \rho_\GC )$ 
at long times;
and depolarization increases both relative entropies. Due to this parallel increase, and because the experimental noise appears to be weak, noise is again expected not to affect our conclusion: Regardless of noise, $\rho_\GC$ should predict the long-time state best under the $XY$ model's evolution.

\section{Distinction between non-Abelian thermal state and competitors at all global system sizes}
\label{app_Cirac_Arg} 

The main text answers a question established in~\cite{NYH_20_Noncommuting}:
Consider a global system of $\Sites$ subsystems,
which exchange noncommuting charges.
Consider measuring one subsystem's long-time state.
Measure the state's distance to $\rho_\NATS$
and to the competitor thermal states:
the canonical $\rho_\can$ and
the grand canonical $\rho_\GC$.
The NATS was found numerically, in~\cite{NYH_20_Noncommuting}, 
to predict the final state most accurately.
However, as $\Sites$ grew, $\rho_\can$ and $\rho_\GC$
approached $\rho_\NATS$ in accuracy.
The reason was believed to be the initial global state,
which had a high temperature and low chemical potentials (see 
Sec.~\ref{sec_Methods}).
Does $\rho_\NATS$ remain substantially more accurate
at all $\Sites$, for any initial state $\ket{\psi_0}$?
Or do all the thermal states' predictions converge
in the thermodynamic limit (as $\Sites \to \infty$), 
for every $\ket{\psi_0}$?

Our experiment suggests the former, as explained in Sec.~\ref{sec_Results}.
We constructed a $\ket{\psi_0}$ for which
the NATS prediction remains more accurate
than the $\rho_\can$ and $\rho_\GC$ predictions,
by approximately constant-in-$\Sites$ amounts,
at all $\Sites$ values realized experimentally.
(Fig.~\ref{fig_Drel_to_exp_NrQubits}).
Section~\ref{sec_Methods} provides one perspective on why this $\ket{\psi_0}$ 
distinguishes the thermal states.
We provide another perspective here.

We prove that, under conditions realized in our experiment,
$\rho_\NATS$, averaged over space, differs from
the average $\rho_\can$ and $\rho_\GC$.
This difference remains nonzero even in the thermodynamic limit.
Appendix~\ref{sec_CiracArg_Setup} presents the setup,
which generalizes our experiment's.
In App.~\ref{sec_CiracArg_Results}, we formalise and prove the result.\footnote{
We thank Ignacio Cirac for framing this argument.}
Appendix~\ref{sec_CiracArg_Exp_Check} shows how our experiment realizes the general setup.

\subsection{Setup}
\label{sec_CiracArg_Setup}

Consider a global system of $\Sites$ identical subsystems.
Let $O^\JParen$ denote observable $O$ of subsystem $j$.
Sometimes, $O^\JParen$ will implicitly be padded with 
identity operators $\id$ acting on the other subsystems.
The corresponding global observable is
$O^\tot  :=  \sum_{j = 1}^\Sites  O^\JParen$.

The Hamiltonian $H^\tot$ is translationally invariant.
$H^\tot$ conserves global charges 
$Q_\gamma^\tot
:=  \sum_{j = 1}^\Sites  Q_\gamma^\JParen$,
for $\gamma = 0, 1, 2, \ldots, c$.
The charges do not all commute pairwise:
$[Q_\gamma^\JParen,  Q_{\gamma'}^\JParen]  \neq  0$ 
for at least one pair $(\gamma, \gamma')$.

We assume that some global unitary $V$ satisfies two requirements.
First, the unitary commutes with the Hamiltonian: $[V, H^\tot] = 0$.
Second, conjugating at least one global charge $Q_\gamma^\tot$ with $U$ 
negates the charge: 
\begin{align}
   \label{eq_U_Q}
   V Q_\gamma^\tot  V^\dag  =  - Q_\gamma^\tot .
\end{align}
We assume that this global charge's initial expectation value
is proportional to the global system size, as in the trapped-ion experiment:
\begin{align}
   \label{eq_Init_Q_Nonzero}
   \expval{ Q_\gamma^\tot }_0
   = q_\gamma \Sites  
   \neq  0 ,
\end{align}
for some constant-in-$\Sites$ $q_\gamma$.

Let $\ket{\psi_0}$ denote the initial global state.
It is invariant, we assume, under translations through
$\trans$ sites, for some non-negative integer $\trans$.
More precisely, divide the chain into clumps of $\trans$ subsystems.
Index the clumps with $m = 1, 2, \ldots, \Sites / \trans$.
(We assume for convenience that $\Sites$ is 
an integer multiple of $\trans$.)
Consider tracing out all the subsystems except the $m^\th$ clump:
$\Tr_{ 1, 2, \ldots, (m-1) \tau,  \; \;  m \tau + 1, m \tau + 2, \ldots, \Sites } 
   (  \ketbra{\psi_0}{\psi_0}  ) .$
This state's form does not depend on $m$.

Let us define a state averaged over clumps of subsystems. 
Let $\rho$ denote any state of the global system.
Consider the clump that, starting at subsystem $j$, 
encompasses $\trans$ subsystems.
This clump occupies the state
\begin{align}
   \label{eq_One_Clump_t}
   \rho^{(j, j+1, \ldots, j + \trans - 1)}
   := \Tr_{1, 2, \ldots, j - 1,  \; \;  j + \trans, j + \trans + 1, \ldots, \Sites}
   (\rho) .
\end{align}
Let $\Trans_j$ denote the operator that translates 
a state $j - 1$ sites leftward. We define the average
\begin{align}
   \label{eq_Rho_Avg}
   \rho^\avg
   :=  \frac{1}{\Sites}  \sum_{j = 1}^{\Sites} 
   \Trans_j  \left(
   \rho^{(j, j\oplus1, \ldots, j \oplus \trans \ominus 1)}  \right)
\end{align}
on the joint Hilbert space of subsystems 1 through $\trans$.
Addition and subtraction modulo $\Sites$ are denoted by $\oplus$ and $\ominus$.
If $\rho$ is fully translationally invariant (if $\trans =1$),
then $\rho^\avg = \Tr_{2, 3, \ldots, \Sites} ( \rho )$,
and this definitional step can be skipped.

Multiple thermal states will interest us.
The global canonical state is defined as 
$\rho_\can^\tot
   :=  \exp( - \beta H^\tot ) / Z_\can^\tot .$
The partition function is 
$Z_\can^\tot := \Tr ( e^{- \beta H^\tot} )$. 
The inverse temperature $\beta$ is defined through
$\bra{\psi_0} H^\tot \ket{\psi_0}
   = \Tr  \left(  H^\tot  \rho_\can^\tot  \right) .$
Define the single-site 
$\rho_\can^\JParen := \Tr_{ \bar{j} } ( \rho_\can^\tot )$.
Denote by $\rho^\avg_\can$ the result of 
averaging $\rho^\tot_\can$ over clumps, as in Eq.~\eqref{eq_Rho_Avg}.

The global NATS is defined as 
\begin{align}
   \label{eq_Def_NATS}
   \rho_\NATS^\tot
   :=  \exp \left( - \beta \left[ H^\tot 
   - \sum_{\gamma = 1}^c  \mu_\gamma  Q_\gamma^\tot  
   \right]  \right) / Z_\NATS^\tot .
\end{align}
This $\beta$ is defined analogously to the canonical $\beta$.
The temperatures' values might differ, 
but we reuse the symbol $\beta$ for convenience.
The effective chemical potentials $\mu_\gamma$ are defined through~\cite{NYH_20_Noncommuting}
\begin{align}
   \label{eq_mu}
   q_\gamma  \Sites
   = \Tr \left(  Q_\gamma^\tot   \rho_\NATS^\tot  \right) .
\end{align}
The partition function 
$Z_\NATS^\tot := \Tr \left(  
  e^{- \beta \left( H^\tot 
   - \sum_{\gamma = 1}^c  \mu_\gamma  Q_\gamma^\tot  
   \right) }  \right)$.
Define $\rho_\NATS^\JParen$ and $\rho^\avg_\NATS$
analogously to $\rho_\can^\JParen$ and $\rho_\can^\avg$.

Our argument concerns multiple distance measures.
Let $O$ denote an arbitrary observable defined on 
an arbitrary Hilbert space.
The Schatten $p$-norm of $O$ is
$|| O ||_p
:=  \left[  \Tr \left(  | O |^p  \right)  \right]^{1/p}$, wherein
$| O |  :=  \sqrt{ O^\dag O }$ and $p \in [0, \infty)$.
The limit as $p \to \infty$ yields the operator norm:
$\lim_{p \to \infty} || O ||_p
=: || O ||_{\rm op}$.
Let $\rho$ and $\sigma$ denote operators
defined on an arbitrary Hilbert space.
The Schatten $p$-distance between the states is
$|| \rho - \sigma ||_p$. 
The trace distance is $\dist (\rho, \sigma) = \frac{1}{2} || \rho - \sigma ||_1$.

\subsection{Lower bounds on distances between thermal states}
\label{sec_CiracArg_Results}

We now formalise the result.

\begin{theorem}
Let the setup and definitions be as in the previous subsection.
Consider the distance from the average NATS 
to the average canonical state.
Measured with the Schatten 1-distance or the relative entropy,
this distance obeys the lower bound
\begin{align}
   \label{eq_NATS_to_Can_1norm}  
   D \left(  \rho_\NATS^\avg  ||  \rho_\can^\avg  \right)
   \geq   \dist  \left(
   \rho_\NATS^\avg,  \rho_\can^\avg
   \right)
   \geq  \frac{ | q_\gamma | }{
                \left\lvert  \left\lvert Q_\gamma^\JParen 
                \right\rvert  \right\rvert_{\rm op} }
   > 0   \, ,
\end{align}
for an arbitrary $j = 1, 2, \ldots, \Sites$.
$\rho_\can$ can be replaced with any grand canonical state that 
commutes with $V$.
\end{theorem}
\noindent
The bound does not depend on $\Sites$
and so holds in the thermodynamic limit.

\begin{proof}

The proof has the following outline. 
First, we calculate the expectation value of $Q_\gamma^\1$
in $\rho_\NATS^\avg$; the result is $q_\gamma$.
Second, the expectation value in $\rho_\can^\avg$ vanishes,
we show using $V$.
Because the two expectation values differ,
a nonzero Schatten 1-distance separates the states.
The Schatten 1-distance lower-bounds the relative entropy
via Pinsker's inequality.

$Q_\gamma^\1$ has an expectation value, in the average NATS state, of
\begin{widetext}
\begin{align}
   \label{eq_NATS_Expval1}
   \Tr \LParen  Q_\gamma^\1  \rho^\avg_\NATS  \RParen
   & = \frac{1}{\Sites}  \sum_{j=1}^\Sites
   \Tr  \left(  Q_\gamma^\1 \,
   \Trans_j \left(  \rho_\NATS^{(j, j \oplus 1, \ldots, j \oplus \trans \ominus 1)}  \right)  \right) \\
   \label{eq_NATS_Expval1b}
   & =  \frac{1}{\Sites}  \sum_{j=1}^\Sites
   \Tr  \left(  Q_\gamma^\JParen  \,
   \rho_\NATS^\JParen  \right)  \\
   & =  \frac{1}{\Sites}  \Tr  \left(
   \left[  \sum_{j=1}^\Sites  
            \id^{\otimes (j - 1)}  \otimes
            Q_\gamma^\JParen  \otimes
            \id^{\otimes (\Sites - j)}   \right]
   \left[  \bigotimes_{k = 1}^\Sites
            \rho_\NATS^{(k)}   \right]
   \right)  \\
   & =  \frac{1}{\Sites}  \Tr  \left(
   Q_\gamma^\tot  \,  
   \rho_\NATS^\tot  \right) \\
   \label{eq_NATS_Expval2}
   & =  q_\gamma .
\end{align}
\end{widetext}
Equation~\eqref{eq_NATS_Expval1b} follows from the definition of $\Trans_j$.
Equation~\eqref{eq_NATS_Expval2} follows Eq.~\eqref{eq_mu}.

The analogous canonical expectation value vanishes, we show next.
We begin with the global expectation value
$\Tr ( Q_\gamma^\tot  \,  e^{- \beta H^\tot} ) / Z_\can^\tot$.
By Eq.~\eqref{eq_U_Q}, we can replace the
$Q_\gamma^\tot$ with $- V Q_\gamma^\tot  V^\dag$.
We then invoke the trace's cyclicality:
\begin{align}
   \label{eq_Can_Expval}
   \Tr  \left( Q_\gamma^\tot  \,  e^{- \beta H^\tot} 
   \right) / Z_\can^\tot
   & =  - \Tr  \left(  
   \left[ V Q_\gamma^\tot V^\dag \right]  \, 
   e^{ - \beta H^\tot }  \right)  /  Z_\can^\tot \\
   & =  - \Tr \left(  Q_\gamma^\tot
   \left[  V^\dag  e^{ - \beta H^\tot }  V  \right]
   \right)  /  Z_\can^\tot \\
   \label{eq_Can_Expval2}
   & = - \Tr \left(  Q_\gamma^\tot
  e^{ - \beta H^\tot }   \right)  /  Z_\can^\tot .
\end{align}
Equation~\eqref{eq_Can_Expval2} follows from $[V, H^\tot] = 0$.
Let us compare the beginning and end of 
Eqs.~\eqref{eq_Can_Expval}--\eqref{eq_Can_Expval2}.
The expectation value 
$\Tr  \left( Q_\gamma^\tot  \, 
e^{ - \beta H^\tot }   \right)  /  Z_\can^\tot$
equals its negative and so vanishes.
We can re-express the null expectation value in terms of
the average canonical state:
\begin{align}
   0
   \label{eq_Can_Help0}
   & = \sum_{j = 1}^\Sites  \Tr \left(
   Q_\gamma^\tot   e^{ - \beta H^\tot }   \right)  /  Z_\can^\tot \\
   & = \sum_{j = 1}^\Sites  \Tr \left(
   \left[ \id^{\otimes (j-1)}  \otimes
   Q_\gamma^\JParen  \otimes
   \id^{\otimes (\Sites - j) }  \right]
   e^{- \beta H^\tot}  \right) /  Z_\can^\tot  \\
   & = \sum_{j = 1}^\Sites  \Tr \left(
   Q_\gamma^\JParen
   \Tr_{\bar{j}}  \left(  e^{- \beta H^\tot}  \right) 
   \right) / Z_\can^\tot \\
   \label{eq_Can_Help1}
   & = \sum_{j = 1}^\Sites  \Tr_j
   \left( Q_\gamma^\JParen  \rho_\can^\JParen  \right) \\
   \label{eq_Can_Help2}
   & = \sum_{j = 1}^\Sites  \Tr  
   \left( Q_\gamma^\1   \Trans_j  
   \left(  \rho_\can^{(j, j \oplus 1, \ldots, j \oplus \tau \ominus 1)}  \right)  \right) \\
   \label{eq_Can_Help3}
   & = \Tr \left(  Q_\gamma^\1  \rho_\can^\avg  \right) .
\end{align}
Equations~\eqref{eq_Can_Help1} and~\eqref{eq_Can_Help2}
are analogous to Eqs.~\eqref{eq_NATS_Expval1b} and~\eqref{eq_NATS_Expval1}.

We have calculated two expectation values of $Q_\gamma^1$,
one in $\rho_\NATS^\avg$ and one in $\rho_\can^\avg$.
The two expectation values differ, by Eqs.~\eqref{eq_Can_Help0},~\eqref{eq_Can_Help3}, 
and~\eqref{eq_NATS_Expval2}:
\begin{align}
   \label{eq_Diff_Expvals}
   \left\lvert
   \Tr  \left(  Q_\gamma^\1  \rho^\avg_\NATS  \right)
   -  \Tr  \left(  Q_\gamma^\1  \rho_\can^\avg  \right)
   \right\rvert
   =  | q_\gamma |
   > 0 .
\end{align}

The absolute difference~\eqref{eq_Diff_Expvals},
we can relate to the trace distance. 
Let $\rho$ and $\sigma$ denote quantum states
defined on an arbitrary Hilbert space.
The interstate distance equals a supremum over 
observables $O$ defined on the same space~\cite[Lemma~9.1.1]{Wilde_11_From}:
\begin{align}
   \label{eq_Sup_Bound}
   \dist (\rho, \sigma)
   & =  \sup_{O \, : \,  || O ||_{\rm op}  \leq 1}
   \Set{ | \Tr (\rho O)  -  \Tr (\sigma O) | } \, .
\end{align}
Let $\rho = \rho_\NATS^\avg$ and
$\sigma = \rho_\can^\avg$.
The operator $Q_\gamma^\1 / || Q_\gamma^\1 ||_{\rm op}$ 
is one normalized $O$. 
Therefore, by Eq.~\eqref{eq_Diff_Expvals},
$\frac{ | q_\gamma | }{ || Q_\gamma^\1 ||_{\rm op}  }$
lower-bounds the supremum in~\eqref{eq_Sup_Bound}.
The superscript $(1)$ can be replaced with $(j)$,
due to translation invariance in the $\Trans_j$ argument. Hence
$\dist (  \rho^\avg_\NATS,  \rho_\can^\avg  )
   \geq  \frac{ | q_\gamma | }{
                 || Q_\gamma^\JParen ||_{\rm op} }
   >  0 \, .$
The final inequality follows from (i) the assumption~\eqref{eq_Init_Q_Nonzero}
and (ii) the finiteness of the single-subsystem 
$|| Q_\gamma^\JParen ||_{\rm op}$.
The first inequality in~\eqref{eq_NATS_to_Can_1norm} follows
via Pinsker's inequality: For states $\rho$ and $\sigma$,
$D (\rho || \sigma)  \geq  \dist (\rho,  \sigma)$.
This proof remains true if $\rho_\GC$ replaces $\rho_\can$
and $[\rho_\GC, V] = 0$.

\end{proof}

\subsection{Realization in trapped-ion experiment}
\label{sec_CiracArg_Exp_Check} 

The general setup of App.~\ref{sec_CiracArg_Setup}
can be realized in the main text's trapped-ion experiment.
In the simplest realization, $Q_\gamma = \sigma_x$.
The unitary $V = \sigma_z^{\otimes \Sites}$:
\begin{align}
   V  \sigma_x^\tot  V^\dag
   =  \sigma_z^{\otimes \Sites}
   \left(  \sum_{j = 1}^\Sites  \sigma_x^\JParen  \right)
   \sigma_z^{\otimes \Sites}
   =  \sum_{j = 1}^\Sites  
   \left(  - \sigma_x^\JParen  \right)
   = - \sigma_x^\tot .
\end{align}
The initial state is $\ket{\psi_0} = \ket{x+, y+, z+}^{\otimes \Sites / 3}$,
so $\bra{\psi_0} \sigma_x^\tot  \ket{\psi_0}  \propto  \Sites$,
and the state is invariant under translations through $\trans = 3$ sites.
Define $\rho_\GC^\tot
:= \exp \left( - \beta \left[ H^\tot  -  \mu_z  S_z^\tot  \right]  \right)
/ Z_\GC^\tot$.
The effective chemical potential $\mu_z$ 
is defined as in the main text,
and $Z_\GC^\tot$ normalizes the state.
$\rho_\GC^\tot$ can replace the canonical state in Ineq.~\eqref{eq_NATS_to_Can_1norm}.

The mapping just described is conceptually simple.
However, we find analytically, another mapping achieves 
the tightest bound~\eqref{eq_NATS_to_Can_1norm}:
$\frac{1}{ \sqrt{3} } \left( \sigma_x + \sigma_y + \sigma_z \right)$, and
$V = \left[ \frac{1}{ \sqrt{6} } ( 2 \sigma_x - \sigma_y - \sigma_z )  
\right]^{\otimes \Sites}$.
(Alternatively, the $\sigma_\gamma$'s in $V$ can be permuted in any way.)

\section{Spatiotemporal fluctuations in states' distances to the non-Abelian thermal state}
\label{app_Thermd}

\begin{figure}
    \centering
    \includegraphics[width=88mm]{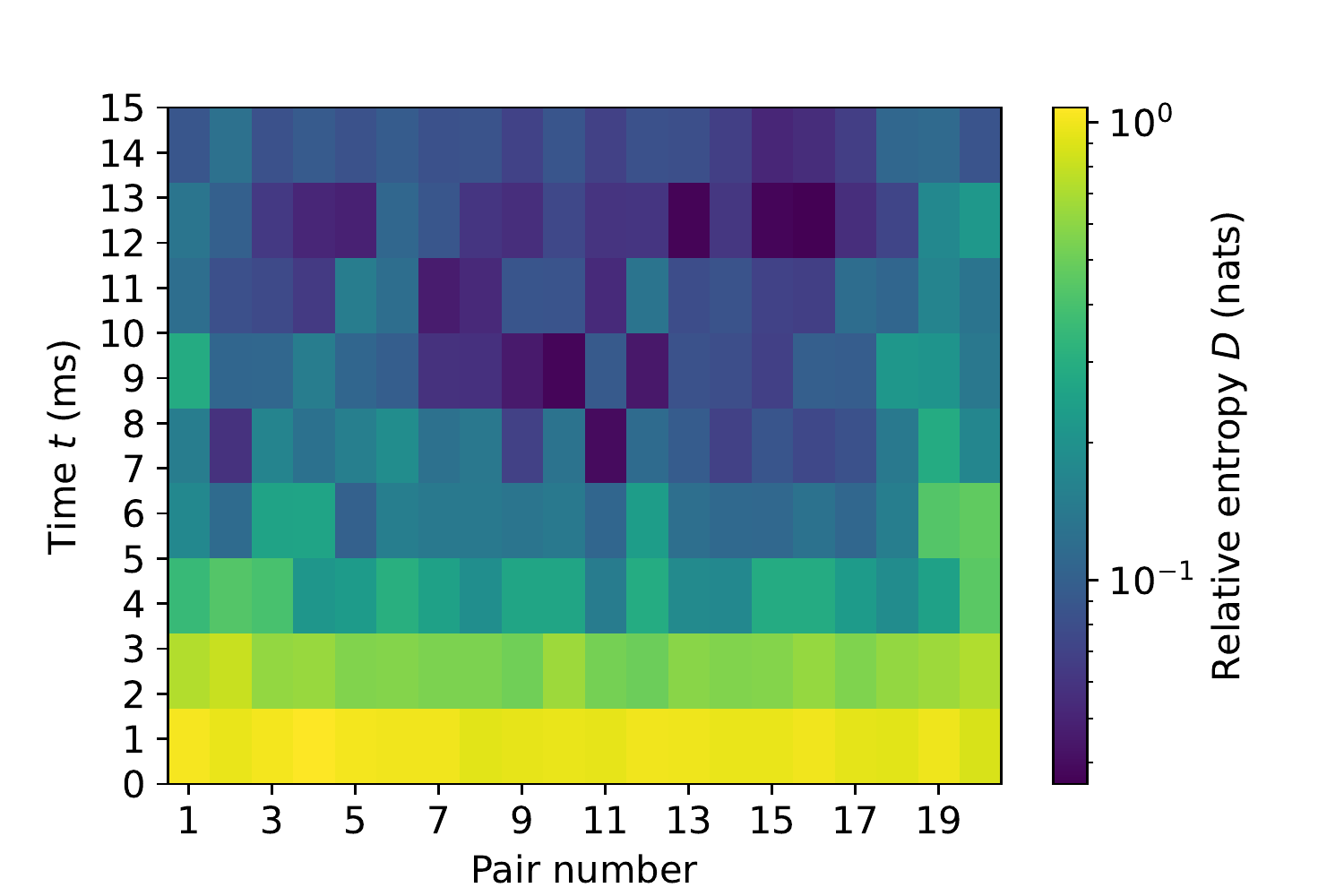}
    \caption{\caphead{Relative entropy to the NATS for each qubit pair, as a function of time:} 
    The spatiotemporal fluctuations show that different qubit pairs thermalize to different extents. The chain consists of $\Sites = 21$ ions.}
    \label{fig_Drel_NATS_pairs}
\end{figure}

Figure~\ref{fig_Drel_NATS_pairs} shows the experimentally observed fluctuations, across space and time, of the relative entropy to the NATS. The chain consists of $\Sites=21$ ions. Every ion pair's state approaches the NATS in time. However, nonuniformity remains; edge pairs thermalize more slowly due to edge effects, while the central pairs thermalize more quickly.

\section{Derivations of Trotter sequences}
\label{app_Trotter}

The evolution implemented differs from evolution under the Heisenberg Hamiltonian~\eqref{eq_Heis} for three reasons.
First, the Heisenberg Hamiltonian is Trotter-approximated.
Second, parts of the Trotter approximation are simulated
via native interactions dressed with rotations.
Third, we reduce decoherence via dynamical decoupling.
Here, we derive the experimental pulse sequence.
We review parts of the setup and introduce notation
in App.~\ref{app_Trotter_Setup}.
In App.~\ref{app_Trotter_Errors}, we detail the two errors against which the pulse sequence protects.
We derive the pulse sequence in App.~\ref{app_Trotter_Derive}.
Appendix~\ref{app_Trotter_XY} extends the derivation 
from the Heisenberg evolution
to the $XY$ model~\eqref{eq_XY_Hamiltonian}.

\subsection{Quick review of setup and notation}
\label{app_Trotter_Setup}

We break a length-$t$ time interval into $\numSteps$ steps
of length $t / \numSteps  =:  \Delta t$ each.
We aim to simulate the Heisenberg Hamiltonian~\eqref{eq_Heis},
whose $J_0 / |j-k|^\alpha$ we sometimes denote by $J_{j,k}$ here.
$H_\Heis$ generates the family of unitaries
$U_\Heis (t)  :=  e^{-i H_\Heis t}$.
To effect this family, we leverage single-axis Hamiltonians
\begin{align}
   \label{eq_H_aa}
   H_{\gamma \gamma}  
   :=  \sum_{j = 1}^\Sites  \sum_{k > j}
   J_{j, k}  \,  \sigma_\gamma^\JParen  \sigma_\gamma^\KParen .
\end{align}
$H_{xx}$ and $H_{yy}$ are native to the experimental platform.
The Hamiltonians~\eqref{eq_H_aa} generate the unitaries
$e^{-i H_{\gamma \gamma} \Delta t} =: U_{\gamma \gamma}$.
We interleave the interaction with rotations 
$R_{\gamma}  
:=  \exp \left(  - i  \,  \frac{\pi}{4}  \,  \sigma_\gamma^\tot  \right)$,
for $\gamma = x, y, z$.
We denote the single-qubit identity operator by $\id$.

\subsection{Two sources of error}
\label{app_Trotter_Errors}

Our pulse sequence combats detuning and rotation errors.
The detuning error manifests as an undesired term
that creeps into the Hamiltonian~\eqref{eq_Heis}.
Proportional to $\sigma_z^\tot$,
the term represents an external magnetic field.
We protect against the detuning error with dynamical decoupling:
The detuning error undesirably rotates
each ion's state about the $z$-axis.
We apply a $\pi$-pulse about the $x$-axis,
reflecting the state through the $xy$-plane.
The state then precesses about the $z$-axis oppositely,
undoing the earlier precession.
Another $\pi$-pulse undoes the reflection.

The second error plagues the engineered rotations:
A qubit may rotate too little or too much, 
because the ion string is illuminated not quite uniformly.
We therefore replace certain rotations $R_\gamma$ with $R_\gamma^\dag$'s.
An ion may rotate too much while undergoing $R_\gamma$ but,
while undergoing $R_\gamma^\dag$,
rotates through the same angle oppositely.
The excess rotations cancel.

\subsection{Derivation of Trotter sequence}
\label{app_Trotter_Derive}

First, we divvy up the Heisenberg evolution into steps.
Then, we introduce rotations that enable dynamical decoupling.
We Trotter-approximate a Heisenberg step in two ways.
Alternating the two Trotter approximations
across a pulse sequence mitigates rotation errors. Engineering of robust Hamiltonians has recently been demonstrated for analog simulations~\cite{Choi_20_Robust, morong2022engineering} and digital circuits~\cite{zhang2022hidden}.

To simulate the Heisenberg Hamiltonian for a time $t$,
we evolve for $\numSteps$ length-$\Delta t$ time steps:
$U_\Heis (t)  =  [ U_\Heis (\Delta t) ]^\numSteps$.
To facilitate dynamical decoupling, 
we insert an identity operator on the left:
$U_\Heis (t) = \id^{\otimes \Sites}  [ U_\Heis (\Delta t) ]^\numSteps$.
We decompose the $\id^{\otimes \Sites}$ into 
rotations about the $z$-axis.
How this decomposition facilitates dynamical decoupling 
is not yet obvious,
as the rotations commute with the detuning expression.
Later, though, we will commute some of the rotations across interaction unitaries.
The commutation will transform the $z$-rotations into $R_x$'s. 
For now, we decompose the $\id^{\otimes \Sites}$ in two ways:
\begin{align}
   \label{eq_Derive_Trotter1}
   U_\Heis (t)  
   & =  (R_z^\dag )^\numSteps  (R_z)^\numSteps
   [ U_\Heis (\Delta t) ]^\numSteps \\
   \label{eq_Derive_Trotter1b}
   & =  (R_z)^\numSteps  ( R_z^\dag )^\numSteps
   [ U_\Heis (\Delta t) ]^\numSteps \ .
\end{align}
We will implement the right-hand side of~\eqref{eq_Derive_Trotter1}
during half the protocol
and, during the other half, implement~\eqref{eq_Derive_Trotter1b}.
This alternation will mitigate rotation errors.

Let us analyze~\eqref{eq_Derive_Trotter1}, then~\eqref{eq_Derive_Trotter1b}.
$R_z$ commutes with $U_\Heis (\Delta t)$ because
the Heisenberg Hamiltonian conserves $\sigma_z^\tot$:
$[ H_\Heis,  \sigma_z^\tot ]  =  0$ implies that
$[ U_\Heis(\Delta t),  R_z]  =  0 .$
The $R_z$'s of Eq.~\eqref{eq_Derive_Trotter1} can therefore move
inside the square brackets:
\begin{align}
   \label{eq_Derive_Trotter2}
   U_\Heis (t)  
   =  (R_z^\dag )^\numSteps
   \left[  R_z  U_\Heis (\Delta t)  \right]^\numSteps \ .
\end{align}
We Trotter-approximate the short Heisenberg evolution as 
\begin{align}
   \label{eq_Derive_Trotter3}
   U_\Heis (\Delta t)
   \approx  U_{yy}  U_{zz}  U_{xx} .
\end{align}
The ordering of the directions is arbitrary.

We substitute into~\eqref{eq_Derive_Trotter2}
and rewrite the bracketed factor, pursuing three goals.
First, the $U_{zz}$ is not native to our platform.
We therefore simulate it with $R_y^\dag U_{xx} R_y$.
Second, one $R_x$ must end up amidst the $U_{\gamma \gamma}$'s.
Two blocks of $U_{\gamma \gamma}$'s, each containing an $R_x$,
will consequently effect one $\pi$ pulse.
Composing these $\pi$ pulses will effect dynamical decoupling.
Third, any other, stray $R_\gamma$'s must be arranged symmetrically
on either side of the $U_{\gamma \gamma}$'s, as explained below.

Let us replace the $U_{zz}$ in~\eqref{eq_Derive_Trotter3} with $R_y^\dag U_{xx} R_y$.\footnote{
One can prove the expressions' equality by writing out
the Taylor series for $U_{xx} = \exp ( i \Delta t H_{xx} )$,
conjugating each term with the rotations,
invoking the Euler decomposition
$R_y  =  \cos (\pi / 4) \id^{\otimes \Sites}  
-  i \sin (\pi / 4) \sigma_y^\tot$, 
multiplying out, and invoking 
$\sigma_\lambda \sigma_\nu
= i \varepsilon_{\lambda \nu \xi} \sigma_\xi$.
The result is the Taylor series for $U_{zz}$.}
The $R_y^\dag$ commutes across the $U_{yy}$:
\begin{align}
   \label{eq_Derive_Trotter4}
   R_z  U_{yy}  U_{zz}  U_{xx}
   & =  R_z  U_{yy}  ( R_y^\dag  U_{xx}  R_y )  U_{xx} \\
   & =  R_z  R_y^\dag  U_{yy}  U_{xx}  R_y  U_{xx} .
\end{align}
We have eliminated the $U_{zz}$.
Similarly eliminating the $R_z$ will prove useful,
so we invoke $R_z = R_y^\dag  R_x  R_y$:
\begin{align}
   R_z  U_{yy}  U_{zz}  U_{xx}
   \label{eq_Derive_Trotter5}
   & =  (R_y^\dag  R_x  \cancel{R_y} ) 
   \cancel{R_y^\dag}  U_{yy}  U_{xx}  R_y  U_{xx} \\
   & =   (R_y^\dag  R_x)  U_{yy}  U_{xx}  R_y  U_{xx}.
\end{align}
We will benefit from complementing the $R_y^\dag  R_x$ 
with a mirror image $(R_y^\dag R_x)^\dag  =  R_x^\dag R_y$
on the right:
We will implement $R_z  U_{yy}  U_{zz}  U_{xx}$ many times,
and instances of the left-hand $R_y^\dag  R_x$ will cancel 
instances of the right-hand $R_x^\dag R_y$.
Therefore, we insert 
$\id^{\otimes \Sites}  
=  R_y^\dag  R_x  R_x^\dag  R_y$
into the right-hand side of~\eqref{eq_Derive_Trotter5}:
\begin{align}
   R_z  U_{yy}  U_{zz}  U_{xx}
   \label{eq_Derive_Trotter6}
   =  (R_y^\dag  R_x)
   U_{yy}  U_{xx}  
   \underbrace{ R_y  U_{xx} (R_y^\dag }_{ \mathrlap{ = U_{zz} } }
   R_x  R_x^\dag  R_y) .
\end{align}
Again, $U_{zz}$ is not native to our platform.
We therefore commute the $R_x$ across the $U_{zz}$,
invoking $R_x^\dag  U_{zz}  R_x  =  U_{yy}$:
\begin{align}
   R_z  U_{yy}  U_{zz}  U_{xx}
   \label{eq_Derive_Trotter8}
   =  (R_y^\dag  R_x)
   \underbrace{ U_{yy}  U_{xx}  R_x  U_{yy} }_{ \mathrlap{ =: E_+ } }
   (R_x^\dag  R_y) .
\end{align}
The final expression has the sought-after form.
We substitute into Eq.~\eqref{eq_Derive_Trotter3},
then into Eq.~\eqref{eq_Derive_Trotter2},
and then cancel rotations:
$U_\Heis (t)
   \approx  (R_z^\dag)^\numSteps  
   (R_y^\dag  R_x)  \,
   ( E_+ )^\numSteps
   (R_x^\dag  R_y) .$

Suppose that $\numSteps = 4$. 
The $E_+$'s, containing four $R_x$'s total,
implement two $\pi$ pulses---one round of dynamical decoupling.
Furthermore, $(R_z^\dag)^4 = (-1)^\numSteps \id^{\otimes \Sites}$, so
\begin{align}
   U_\Heis ( 4 \Delta t )
   & \approx  (R_z^\dag)^4  
   ( R_y^\dag  R_x )  \,
   ( E_+ )^4
   ( R_x^\dag  R_y ) \\
   & =  (-1)^\Sites  
   (R_y^\dag  R_x)  \,
   ( E_+ )^4
   (R_x^\dag  R_y) .
    \label{eq_Derive_Trotter9}
\end{align}

Now, let $\numSteps \gg 4$, as in the experiment.
After one round of dynamical decoupling, 
to mitigate the detuning error,
we mitigate rotation errors.
We effect four time steps with an alternative operator
derived from Eq.~\eqref{eq_Derive_Trotter1b}.
Then, we continue alternating.

Let us derive the alternative to $E_+$. 
We shift the $R_z^\dag$'s of Eq.~\eqref{eq_Derive_Trotter1b}
inside the square brackets:
\begin{align}
   \label{eq_Derive_Trotter10}
   U_\Heis (t)
   & =  (R_z)^\numSteps
   \left[ R_z^\dag  U_\Heis (\Delta t)  \right]^\numSteps  \\
   & \approx  (R_z)^\numSteps
   \left[  R_z^\dag
   U_{yy}  U_{zz}  U_{xx}  \right]^\numSteps \ .
\end{align}
The final expression follows from Eq.~\eqref{eq_Derive_Trotter3}.
The bracketed factor must end up with the 
$(R_y^\dag  R_x)  [ \ldots ]  (R_x^\dag  R_y)$ 
structure of Eq.~\eqref{eq_Derive_Trotter8},
so that rotations cancel between instances of~\eqref{eq_Derive_Trotter8}
and instances of the new bracketed factor.
We therefore ensure that $R_y^\dag R_x$ is on the factor's left-hand side,
then propagate extraneous rotations leftward:
\begin{align}
   & \underbrace{ R_z^\dag }_{ \mathrlap{
      = R_y^\dag  R_x^\dag  R_y
      = R_y^\dag  \id^{\otimes \Sites}  R_x^\dag  R_y
      = R_y^\dag  (R_x  R_x^\dag)  R_x^\dag  R_y } }
   U_{yy}  U_{zz}  U_{xx} \\
   & = ( R_y^\dag  R_x )  (R_x^\dag)^2  
   \underbrace{ R_y  U_{yy} }_{\mathrlap{ = U_{yy} R_y } }  
   U_{zz}  U_{xx}  \\
   & = ( R_y^\dag  R_x )  R_x^\dag
   \underbrace{ R_x^\dag  U_{yy} }_{\mathllap{ = U_{zz} R_x^\dag } }
   \underbrace{ R_y  U_{zz} }_{ \mathrlap{ = U_{xx} R_y } }
   U_{xx} \\
   & = ( R_y^\dag  R_x )  
   \underbrace{ R_x^\dag  U_{zz} }_{ \mathllap{ = U_{yy} R_x^\dag } }
   \underbrace{ R_x^\dag  U_{xx} }_{ = U_{xx}  R_x^\dag }
   \underbrace{ R_y  U_{xx} }_{ \mathrlap{ = U_{zz} R_y } }  \\
   & =  ( R_y^\dag  R_x )  U_{yy} 
   \underbrace{ R_x^\dag  U_{xx} }_{ \mathllap{ = U_{xx} R_x^\dag } }
   \underbrace{ R_x^\dag  U_{zz} }_{ \mathrlap{ = U_{yy} R_x^\dag} }
   R_y \\
   & =  ( R_y^\dag  R_x )  
   \underbrace{ U_{yy}   U_{xx}  R_x^\dag  U_{yy} }_{ \mathrlap{ =: E_- } }
   (R_x^\dag  R_y) .
\end{align}

By Eq.~\eqref{eq_Derive_Trotter10},
${ U_\Heis (t)
   \approx  (R_z)^\numSteps
   (R_y^\dag  R_x)
   (E_-)^\numSteps
   (R_x^\dag  R_y) }$.
Analogously to Eq.~\eqref{eq_Derive_Trotter9},
\begin{align}
   \label{eq_Derive_Trotter12}
   U_\Heis (4 \Delta t) 
   \approx  (-1)^\Sites 
   (R_y^\dag  R_x)
   (E_-)^4
   (R_x^\dag  R_y) .
\end{align}

We alternate instances of~\eqref{eq_Derive_Trotter9} 
with instances of~\eqref{eq_Derive_Trotter12}
to simulate long Heisenberg evolutions.
Many rotations cancel. 
If $\numSteps$ equals an integer multiple of eight,
\begin{align}
    \label{eq:Trotter_Heis_alt}
   U_\Heis (t) 
   \approx  R_y^\dag  R_x
   \left[ (E_-)^4  (E_+)^4  \right]^{\numSteps / 8}
   R_x^\dag  R_y .
\end{align}

\subsection{Extension from Heisenberg model to \texorpdfstring{$XY$}{XY)} model}
\label{app_Trotter_XY}

In the Results, we experimentally compared 
the Heisenberg evolution with 
evolution under the $XY$ model, Eq.~\eqref{eq_XY_Hamiltonian}.
$H_{xy}$ generates the unitaries
$U_{xy} (t)  :=  \exp ( -it H_{xy} )$.
We can more easily Trotterize $U_{xy}(t)$ while mitigating errors than Trotterize $U_\Heis (t)$.

As before, we divvy up the evolution into steps.
Then, we Trotter-approximate the steps and insert 
$\id^{\otimes \Sites} = \left[  \left(  R_x^\dag  \right)^2  \right]^\numSteps
   \left(  R_x^2  \right)^\numSteps$:
\begin{align}
   U_{xy} (t) 
   & =  [ U_{xy} (\Delta t) ]^\numSteps 
   \approx ( U_{yy} U_{xx} )^\numSteps \\
   \label{eq_Trotter_XY_Help1}
   & = \left[  \left(  R_x^\dag  \right)^2  \right]^\numSteps
   \left(  R_x^2  \right)^\numSteps
   ( U_{yy} U_{xx} )^\numSteps \ .
\end{align}
Due to the square, $R_x^2$ commutes with $U_{yy} U_{xx}$:
\begin{align}
   R_x^2  U_{yy} U_{xx}
   & = R_x 
   \underbrace{ R_x  U_{yy} }_{ \mathrlap{= U_{zz} R_x} }
   U_{xx}
   = \underbrace{ R_x  U_{zz} }_{ \mathrlap{= U_{yy}  R_x} } 
   R_x  U_{xx} \\
   & =  U_{yy}  
   \underbrace{ R_x^2  U_{xx} }_{ \mathrlap{= U_{xx}  R_x^2} }
   =  U_{yy}  U_{xx}  R_x^2 \ .
\end{align}
Therefore, in Eq.~\eqref{eq_Trotter_XY_Help1},
we can pull the $( R_x^2 )^\numSteps$ into the parentheses:
\begin{align}
   \label{eq_Trotter_XY_Help2}
   U_{xy} (t) 
   =  \left[ \left( R_x^\dag \right)^2  \right]^\numSteps 
   \left(  R_x^2  U_{yy}  U_{xx} \right)^\numSteps .
\end{align}
We could commute the $R_x^2$ into the center of 
the $U_{\gamma \gamma}$'s,
to improve the dynamical decoupling.
However, Eq.~\eqref{eq_Trotter_XY_Help2} suffices;
errors accumulate in only a couple of gates.

The operator $F_+ := R_x^2  U_{yy}  U_{xx}$ contains a $\pi$-pulse.
Therefore, we need perform $F_+$ only twice
before implementing $F_- :=  (R_x^\dag)^2$.
Furthermore, $\left[ \left( R_x^\dag \right)^2  \right]^2  =  (-1)^\Sites$.
If $\numSteps$ is a multiple of four, then
$U_{xy} (t) 
   =  \left[  ( F_- )^2  (F_+)^2  \right]^{\numSteps / 4} \ .$

\section{Assessment of noise-robust Trotter sequence}
\label{app:NumSimDD}

\begin{figure*}[ht!]
    \centering
    \includegraphics[width=180mm]{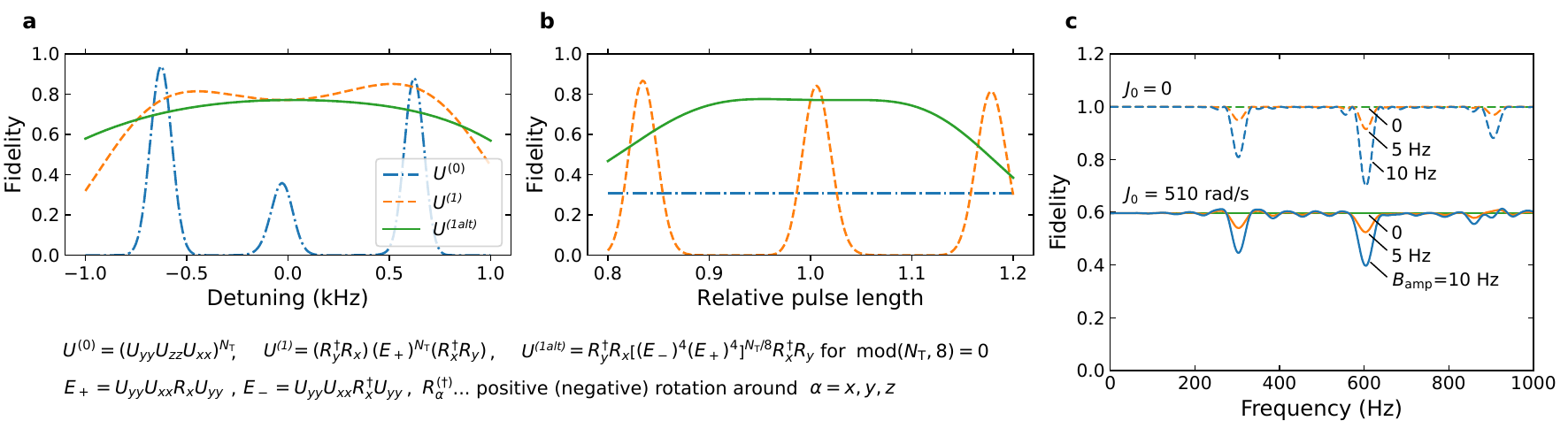}
    \caption{\caphead{Dynamical decoupling:} The simulation was performed with 12 ions, a power-law approximation to the coupling, $J_0=510~\mathrm{rad/s}$, $\alpha=1.02$, and 10~ms of evolution. The fidelity compares the simulated Trotter-approximated state with the exact ideal state: 
    $\left( \Tr \sqrt{ \sqrt{\rho_\mathrm{exact}} \rho_\mathrm{Trotter} \sqrt{\rho_\mathrm{exact}} } \right)^2$. 
    (a)~Introducing $\pi/2$ rotations into the Trotter sequence guards against detuning errors. The right-hand side (RHS) of Eq.~\eqref{eq_Derive_Trotter3} defines the sequence $U^{(0)}$,
    the RHS of Eq.~\eqref{eq_Derive_Trotter12} defines $U^{(1)}$, and the RHS of Eq.~\eqref{eq:Trotter_Heis_alt} defines $U^{(1\mathrm{alt})}$.
    (b)~Alternating the rotation's direction guards against systematic rotation errors. 
    (c)~Response to oscillations of a time-varying magnetic field 
	$\textbf{B}=B_\mathrm{amp} \cos(2\pi f t) \hat z$, wherein $B_\mathrm{amp} \geq 0$ (15~ms evolution). The dynamically decoupled Trotter sequence $U^{(1\mathrm{alt})}$ allows the fidelity to drop.
	The drops occur when the field's frequency, $f$, 
	is an integer multiple of
	$f_1 = \frac{1}{2} (4 T_\mathrm{tot}/N_\mathrm{Trotter})^{-1} 
	= 300~\mathrm{Hz}$.
	We can understand this behavior most simply when $J_0=0$ (top curves): The qubits do not interact, so each qubit remains in a superposition, whose relative phase undesirably changes under~$\textbf{B}$.}
    \label{fig:dynamical_decoupling}
\end{figure*}

Appendix~\ref{app_Trotter} describes the Trotter sequence that we engineered to alleviate errors. Here, we demonstrate the sequence's effectiveness in numerical simulations and in the experiment.
Figure~\ref{fig:dynamical_decoupling} shows the dynamical decoupling's effects in the parameter regime used experimentally. Constant detuning errors of up to several hundred Hertz do not significantly reduce the time-evolved state's fidelity to the ideal state, as shown in panel (a): The fidelity drops by only $<10\%$,
despite detuning errors of up to 500 Hz.
Similarly, systematic rotation errors of $\pm 10\%$ affect the fidelity little [panel (b)]; the fidelity drops by 4\%.
If the detunings oscillate temporally [panel (c)], the dynamical decoupling's robustness depends heavily on the oscillation frequency $f$: Recall that $t_\final = 15$ ms denotes the experiment's temporal length and that $\numSteps$ denotes the number of Trotter steps.
Consider a single-qubit state expressed as a combination of outer products of $\sigma_z$ eigenstates.
If $f$ is an integer multiple of
$f_1 = \frac{1}{2} (4 t_\final / \numSteps )^{-1} 
= 300~\mathrm{Hz}$, the qubit's state acquires a relative phase,
reducing the fidelity to the ideal state.

\begin{figure*}[ht!]
    \centering
    \includegraphics[width=180mm]{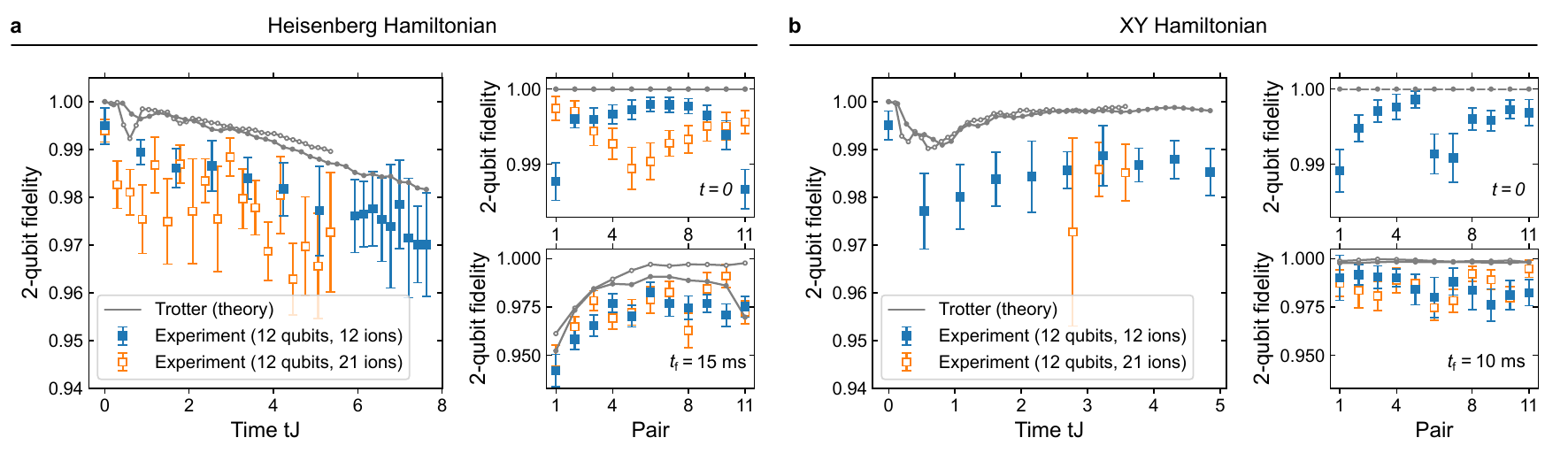}
    \caption{\caphead{Measured fidelity of the state resulting from the Trotter approximation:} 
    The experimentally observed state's fidelity to the ideal state is averaged over ion pairs and plotted against time. Error bars indicate the standard deviation over the pairs. The smaller subplots show the fidelities, at the evolution's start and end, to individual ion pairs' observed states. 
    The two types of markers represent two cases that we compare: First, we realized a 12-qubit system using a chain of only 12 ions (filled blue markers).
    The filled grey dots show the corresponding theoretical prediction.
    Second, we realized a 12-qubit system using a 21-ion chain, by hiding the extraneous ions from the interactions (empty orange markers).
    The open grey dots show this case's theoretical prediction.
    The measurements were carried out (a) for the Trotter-approximated Heisenberg Hamiltonian and (b) for the Trotter-approximated $XY$ Hamiltonian. The fidelity drops and revives at early times. This behavior results from the Trotter steps' failure to conserve the exact Hamiltonian's charges, leading to periodic errors.
    }
    \label{fig:fidelity_evol_time}
\end{figure*}

Figure~\ref{fig:fidelity_evol_time} shows the experimentally observed two-qubit fidelities 
$\left( \Tr \sqrt{ \sqrt{\rho_\mathrm{exact}} \rho_\mathrm{exp} \sqrt{\rho_\mathrm{exact}} } \right)^2$. At $t=0$, the fidelity is limited by imperfections in the state preparation. These imperfections result from the global rotations' inhomogeneous profile (different qubits erroneously rotate by different amounts). Consequently, the initial fidelity is $0.995(4)$, when the Hamiltonian has the Heisenberg form~\eqref{eq_Heis} [Fig.~\ref{fig:fidelity_evol_time}(a)]. 
At $t > 0$, the fidelity is reduced both by Trotterization errors (grey line) and experimental imperfections.
At the final time, $t=t_\final$, the fidelity is $0.97(1)$.

Additionally, we assess the quality of the hiding operation described in 
Sec.~\ref{sec_Methods}: The ion chain always contains 21 ions. However, if we wish to use fewer ions, we hide the extra ions in an extra Zeeman sublevel. To evaluate this technique's effectiveness, we compare two cases: First, we realize a 12-qubit system with a chain of only 12 ions. Second, we realize a 12-qubit system using a 21-ion chain.
Both cases yield similar fidelities in Fig.~\ref{fig:fidelity_evol_time}.
However, the two cases' state-preparation errors differ,
as the preparation requires additional (hiding) operations in the second case.


The $XY$-model Trotterization [Fig.~\ref{fig:fidelity_evol_time}(b)] leads to better fidelities than the Heisenberg-model Trotterization [Fig.~\ref{fig:fidelity_evol_time}(a)].
The reason, we expect, is the $XY$ Trotterization's greater simplicity (requiring fewer steps). At early times, the $XY$-model Trotterization's fidelity drops, then revives. 
This effect is visible for 2-qubit subsystems. It results from finite-length Trotter steps' failure to conserve the exact Hamiltonian's charges. The total system's fidelity drops at all times, numerical simulations (not depicted) show.

\end{appendices}

%
%

\providecommand{\noopsort}[1]{}\providecommand{\singleletter}[1]{#1}%

\end{document}